\documentclass[aps,10pt,groupedaddress]{revtex4-1}
\usepackage[letterpaper,margin=1in]{geometry} 
\usepackage[pdftex]{graphicx}  
\usepackage{color}    
\usepackage[colorlinks=true,citecolor=blue,urlcolor=blue,linkcolor=blue]{hyperref}
\usepackage[font={small},labelfont=bf,margin=10pt,justification=centerlast]{caption}
\usepackage{subcaption}  
\usepackage{hyperref}    
\usepackage[T1]{fontenc} 
\usepackage{enumitem}    
\usepackage{mathtools}
\DeclarePairedDelimiter{\ceil}{\lceil}{\rceil}
\DeclarePairedDelimiter{\floor}{\lfloor}{\rfloor}
\usepackage{yllmath}
\providecommand{\myheading}[1]{\textbf{#1}}

\newcommand{\Ai}{\mathrm{Ai}}
\newcommand{\QB}{1SB}
\newcommand{\SB}{SB}
\newcommand{\PI}{PI}
\newcommand{\tilvi}{u}

\begin{document}
\title{Path-Integral Treatment of Quantum Bouncers}
\author{Yen Lee Loh}
\email{yenlee.loh@und.edu}
\affiliation{Department of Physics and Astrophysics, University of North Dakota, Grand Forks, ND 58202, USA}
\author{Chee Kwan Gan}
\affiliation{Institute of High Performance Computing, 1 Fusionopolis Way, {\#}16-16 Connexis 138632, Singapore}
\date\today  
\begin{abstract}

The one-sided bouncer (\QB{}) and the symmetric bouncer (\SB{}) involve a one-dimensional particle in a piecewise linear potential. 
For such problems, the time-dependent quantum mechanical propagator cannot be found in closed form.  The semiclassical Feynman path integral is a very appealing approach, as it approximates the propagator by a closed-form expression (a sum over a \emph{finite} number of classical paths).
In this paper we solve the classical path enumeration problem.
We obtain closed-form expressions for the initial velocity, bounce times, focal times, action, van Vleck determinant, and Morse index for each classical path. 
We calculate the propagator within the semiclassical approximation.  The numerical results agree with eigenfunction expansion results away from caustics.  
We derive mappings between the \QB{} and \SB{},
which explains why each bounce of the \QB{} increases the Morse index by 2 and results in a phase change of $\pi$.
We interpret the semiclassical Feynman path integral to obtain visualizations of matter wave propagation based on interference between classical paths, in analogy with the traditional visualization of light wave propagation as interference between classical ray paths.
\end{abstract}
\maketitle

\section{Introduction} \label{secIntro}
\begin{figure*}[htbp]
    \begin{subfigure}{.13\textwidth}
        \includegraphics[width=\textwidth]{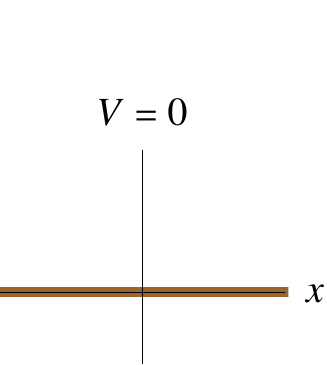}
        \caption{Free particle\label{V1}}
    \end{subfigure}    
    \begin{subfigure}{.13\textwidth}
        \includegraphics[width=\textwidth]{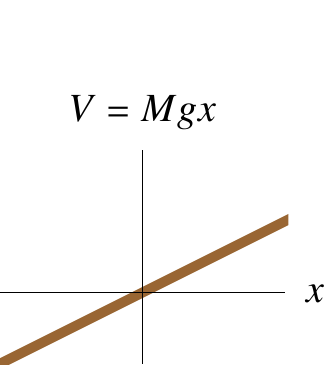}
        \caption{Ramp\label{V2}}
    \end{subfigure}    
    \begin{subfigure}{.13\textwidth}
        \includegraphics[width=\textwidth]{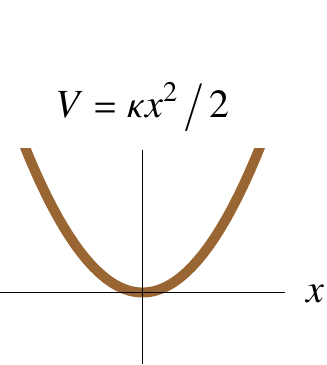}
        \caption{Oscillator\label{V3}}
    \end{subfigure}    
    \begin{subfigure}{.13\textwidth}
        \includegraphics[width=\textwidth]{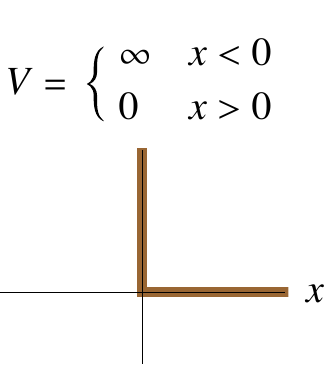}
        \caption{Wall\label{V4}}
    \end{subfigure}    
    \begin{subfigure}{.13\textwidth}
        \includegraphics[width=\textwidth]{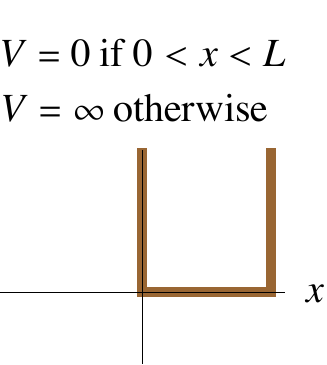}
        \caption{Box\label{V5}}
    \end{subfigure}   
    \begin{subfigure}{.13\textwidth}
        \includegraphics[width=\textwidth]{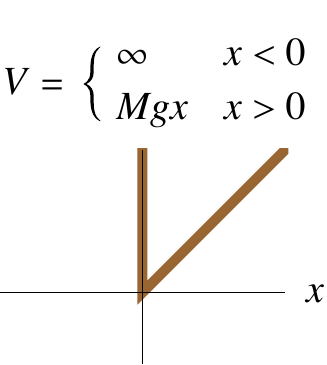}
        \caption{\QB{}\label{VBouncer}}
    \end{subfigure}
    \begin{subfigure}{.13\textwidth}
        \includegraphics[width=\textwidth]{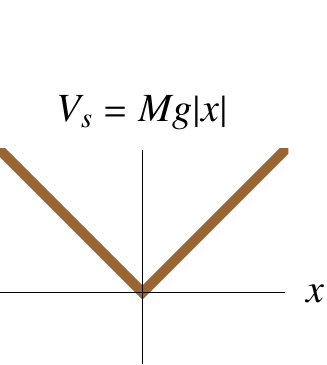}
        \caption{\SB{}\label{VSymmetrized}}
    \end{subfigure}     
    \caption{
	    (a-e) Potentials for problems that are solvable using the \PI{}.
	    (f) Potential for the one-sided bouncer (\QB{}).
	    (g) Potential for the symmetric bouncer (\SB{}).
    }
	\end{figure*}

The Feynman path integral (\PI{}) is a fascinating tool for calculating propagators of quantum-mechanical problems.  It embodies Feynman's view of quantum mechanics \cite{FeynmanLec2005-book} in which a particle takes all possible paths from initial state to final state, weighted by complex phase factors.
It\cite{Feynman2005-book,Sakurai1994-book,Shankar1994-book,Schulman2005-book,Kleinert2006-book} gives complementary insight to eigenfunction expansion methods.
However, the \PI{}\footnote{In this paper, ``PI'' refers exclusively to the real-time Feynman path integral, as opposed to Euclidean or Matsubara path integrals in imaginary time.} is also notorious for numerous mathematical subtleties: 
\begin{enumerate}[noitemsep]
\item It is challenging to define a measure for infinite-dimensional integration.
\item The paths are generally not smooth functions.
\item Na{\"i}ve calculations may be ambiguous with regard to phase factors.\cite{Thornber98v66}
\item Infinite potential barriers pose a problem, because a Gaussian integral cannot be evaluated in closed form if it is restricted to a bounded domain\cite{Goodman81v49}.  Chapter 6 of Schulman's book presents an \emph{ad hoc} prescription, in which Dirichlet boundaries are replaced by images of opposite sign, but this rule was considered to be an ``embarrassment to the purist''\cite{Schulman2005-book}.  A rigorous derivation of the method of images was given by Goodman\cite{Goodman81v49}, but only for the simplest situations.
\end{enumerate}
Thus the \PI{} is often relegated to just being a pedagogical tool, rather than a method for serious calculations.
Indeed, there are only a handful of 1D quantum mechanics problems whose propagators have been calculated using the \PI{}, as depicted in Fig.~\ref{V1}--\ref{V5}.  These include
the free particle (Fig.~\ref{V1});
infinite ramp\cite{Brown94v62,Robinett96v64,Holstein97v65}  (Fig.~\ref{V2});
harmonic oscillator\cite{Schulman2005-book,Chua18v2,Barone03v71,Holstein97v65,Holstein98v66,Moriconi04v72,Shao16v84,Barone03v71,Ponomarenko04v72} (Fig.~\ref{V3});
particle near a wall, a.k.a.~infinite potential barrier \cite{Goodman81v49}
 (Fig.~\ref{V4});
particle in a box, a.k.a.~infinite square well \cite{Goodman81v49}
(Fig.~\ref{V5});
and 
particle on a ring\cite{Schulman2005-book}. 
For cases (a)-(c) the propagator can be calculated by Gaussian fluctuations around a single classical path.  For case (d), one needs to consider two classical paths.  For the particle in a box and particle on a ring, the propagator is a generalized function given by a non-convergent sum over \emph{infinitely many} classical paths, and can neither be evaluated nor plotted.

The one-sided \emph{bouncer} (\QB{}) is a particle on a ramp potential terminated by an infinite potential barrier, as shown in Fig.~\ref{VBouncer}. 
The eigenfunction expansion (EE) for the \QB{} is a well-known textbook problem\cite{Sakurai1994-book,GriffithsQM1994-book,Flugge1994-book}.
It poses greater mathematical complexity than in problems in Figs.~\ref{V1}-\ref{V5}.  
The eigenfunctions are Airy functions instead of exponential and trigonometric functions;
the eigenenergies are incommensurate;
and the propagator cannot be expressed in closed form in terms of named special functions.  
The time-independent Green function of the \QB{} (and similar problems) can be written in closed form in terms of Airy functions.\cite{Glasser15v93}
Whineray\cite{Whineray92v60} gives a matrix method for finding eigenenergies and eigenfunctions.
Wheeler\cite{Wheeler-link} gives a comprehensive overview of the \QB{}, but does not complete the \PI{} analysis.
Goodings\cite{Goodings91v59} finds the semiclassical propagator as a function of energy, extracts the asymptotic eigenenergies and eigenfunctions, and verifies that these agree with the traditional WKB approximation.
Gea-Banacloche\cite{Gea-Banacloche99v67} performed a detailed study of Gaussian wave packet time evolution.

In this paper we treat the \emph{real-time propagator} of the \QB{} and the symmetric bouncer (\SB{}) (Fig.~\ref{VSymmetrized}) using the \PI{}.
We solve the path enumeration problem, obtaining exact closed-form solutions for all classical paths with given end points, for all possible numbers of bounces.  
We determine the phase diagram of the classical bouncer, showing how the number of classical paths depends on the end points and time of flight.
We obtain formulas for the action, van Vleck determinant (VVD), and Morse index (number of foci) for each path.
We calculate the propagator in the semiclassical approximation (SCA) as a sum over all classical paths, where the amplitude of each path depends on the VVD, and the phase of each path depends on the action and the Morse index.
We verify numerically that the SCA agrees well with the EE method in the semiclassical limit (large times and large positions).  
The highlights of our study are:
\begin{enumerate}[noitemsep]
\item Our expressions for the semiclassical propagators are effectively closed-form expressions involving a finite number of standard operations.
\item We present striking visualizations of the \PI{}-SCA approach that show how the quantum phase of a particle evolves along classical paths, in analogy with the textbook diagrams for multiple-slit interference experiments.
\item We prove that the infinite barrier can be eliminated using the method of images, provided that the potential is also symmetrized.  Thus, the propagator of the \QB{} can be found using a subtraction between two values of the propagator of the \SB{}.
\item The \PI{}-SCA gives accurate results for short times and large initial and final positions, where the EE method struggles to converge.
\end{enumerate}
Our work has applications to semiconductors in strong electric fields \cite{Haug2010-book}, neutron beams, atoms and ultracold atomic gases \cite{Desko83v51,Goodings91v59,Whineray92v60,Dembinski93v70,Dembinski96v29,Dowling96v37}, and in fact to any problem involving coherent quantum motion.

This paper is laid out as follows.  We analyze the \QB{} in Sec.~\ref{secBouncer1} and the \SB{} in Sec.~\ref{secBouncer2}.  Note that the calculation of the Morse index is easier for the \SB{}, so the reader may wish to read Sec.~\ref{secMorseIndex2} before Sec.~\ref{secMorseIndex1}.  In Sec.~\ref{secGoodmanSubtraction} we generalize Goodman's path-cancellation argument\cite{Goodman81v49} to show that the method of images can be applied to the \QB{}, thus relating it to the \SB{}.

\section{One-Sided Bouncer}  \label{secBouncer1}
    \begin{figure*}[htbp]
    \begin{subfigure}{.33\textwidth}
        \includegraphics[height=50mm]{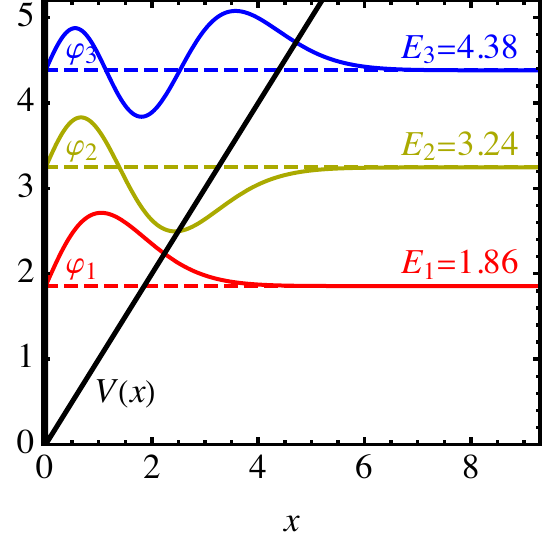}
        \caption{\label{Bouncer1Eigen}}
    \end{subfigure}
    \begin{subfigure}{.64\textwidth}
        \includegraphics[height=50mm]{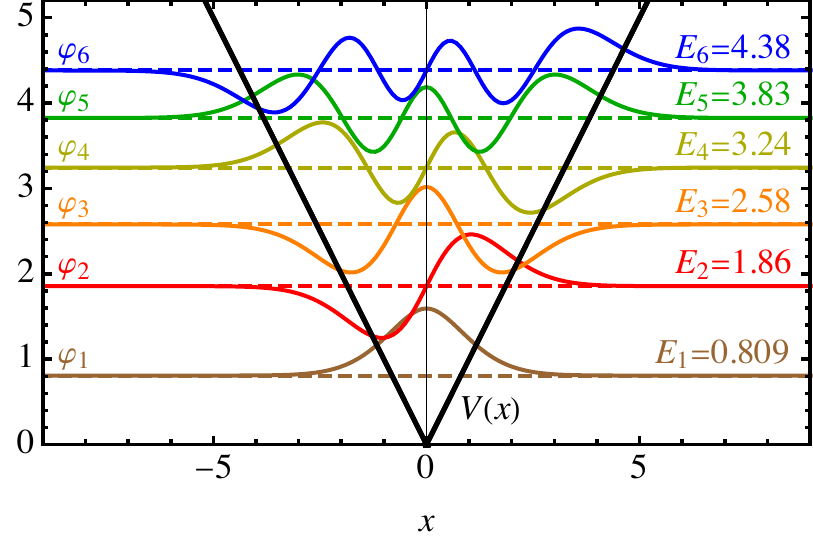}
        \caption{\label{Bouncer2Eigen}}
    \end{subfigure}
    \caption{
	    \label{BouncerEigen}       
	    Eigenenergies $E_n$ and eigenfunctions $\varphi_n(x)$ of the (a) \QB{} and (b) \SB{}.
	    The units for $E_n$ and $x$ are implicit as explained in the text. Note that odd-parity eigenstates of the \SB{} have the same energies as eigenstates of the \QB{}.
    }
	\end{figure*}	

\subsection{Eigenfunction Expansion Method} \label{secEE}
The Hamiltonian for the one-sided bouncer (\QB{})\cite{Gibbs75v43,Gea-Banacloche99v67}
is $\hat{H} = \hat{p}^2 / 2M + V(\hat{x})$, where the potential energy function contains an infinite potential barrier at $x=0$ and a linear gravitational potential energy term:
    \begin{align}
    V(x) &= 
        \begin{cases}
        \infty & x < 0 \\
        M g x  & x > 0.
        \end{cases}
	\label{eqV}
	\end{align}
Here, $M$ is the mass of the particle, $g$ is the uniform gravitational field strength, and $x$ is altitude.
The time-independent Schr\"odinger equation is
$-\frac{\hbar^2}{2M} \frac{d^2\psi}{d x^2} + M g x \psi = E \psi$.
The infinite potential barrier is equivalent to the boundary condition $\psi(0)=0$: the wave function must vanish at the origin.
For this problem it is natural to take the units of energy, length, and time to be 
$E_0 = (\hbar^2g^2M)^{1/3}$, 
$x_0 = (\hbar^2/gM^2)^{1/3}$,
and $T_0 = (\hbar/Mg^2)^{1/3}$, respectively.
(This corresponds to setting $M=g=\hbar=1$.)
Hereafter, for brevity, if a dimensional quantity $Q$ is assigned a numerical value, it is implicitly assumed that $Q$ is expressed in its corresponding basic unit such as $E_0$, $x_0$, or $T_0$.
The eigenenergies and normalized eigenfunctions, illustrated in Fig.~\ref{Bouncer1Eigen}, are
    \begin{align}
    E_n          &= -\lambda_n \frac{E_0}{\gamma}  \\
    \varphi_n(x) &= 
    \left(  \frac{\gamma}{x_0}\right)^{1/2} 
    \frac{\Ai( \frac{\gamma x}{x_0} + \lambda_n)}{\Ai'(\lambda_n)}
    \label{eqEigen}
	\end{align}
where $\gamma=2^{1/3}$.  Here, $\lambda_n$ is the $n$th zero of the Airy function along the negative $x$ axis, such that $\lambda_1 \approx -2.34, \lambda_2 \approx -4.09,$ and so on. 
By expanding the time-dependent Schr\"odinger equation in the eigenfunction basis and taking the initial wave function to be a delta function $\delta(x-x_i)$, one obtains the traditional formula for the propagator as an eigenfunction expansion (EE),
    \begin{align}
    K(x_f, x_i, T)
    &= 
        \sum_{n=1}^\infty
        \varphi_n (x_f) e^{-i E_n T/\hbar} \varphi_n^* (x_i).
    \label{eqKEE}
	\end{align}

Computing the values of Airy zeroes, Airy functions, and Airy derivatives is fairly time-consuming.  Fortunately, Eq.~\eqref{eqKEE} can be written in the form
   \begin{align}
    K(x_f, x_i, T_k)
    &\approx
        \sum_{n=1}^N
        \varphi_{nf} U_{nk} \varphi^*_{ni}    
    \end{align}
where $i$ and $f$ are fixed or variable indices.  The above equation lends itself readily to vectorization.  For example, if $x_i$ is fixed and we wish to evaluate $K$ at final positions $x_f$ ($f=1,2,3,\dotsc,N_x$) and $T=T_k$ ($k=1,2,3,\dotsc,N_T$), we can calculate $K_{fk} = \sum_n \varphi_{nf} U_{nk} \varphi^*_{ni}$ in $O(N_x N_T N)$ time using basic linear algebra operations.

The reader should be warned that for many quantum mechanical problems, the propagator is tremendously ill-behaved.  For example, the propagators of the particle-in-a-box and particle-on-a-ring can be written as sums over eigenstates, which reduce to closed forms containing Jacobi theta functions, but they are unevaluatable and unplottable!\cite{Schulman2005-book})  Such propagators are not ordinary functions, but rather, generalized functions (somewhat like the Dirac delta function itself).  They are mathematical constructs that only reveal their physical meaning when employed as kernels in integrals.

Having said that, our EE results (as well as \PI{} results presented later) demonstrate that the propagators of quantum bouncers are well-behaved.  The sum in Eq.~\eqref{eqKEE} converges to a definite limit, and $K(x_f, x_i, T)$ can be evaluated and plotted.  Convergence is fast at large $T$ and small values of $x_i$ and $x_f$.  However, convergence is slow for small $T$ (e.g., $T<0.4$) and large $x_i$ or $x_f$, even when $10^4$ Airy functions are used in the sum.  This is part of the motivation for developing the \PI{} as an alternative computational method.

\subsection{Feynman Path Integral}  \label{secFPI1}
\myheading{Feynman path integral (\PI{}) formalism.}
For a general one-dimensional quantum Hamiltonian $\hat{H} = \hat{p}^2 / 2M + V(\hat{x})$, the propagator can be written as an \PI{}\cite{Feynman2005-book,Sakurai1994-book,Shankar1994-book,Schulman2005-book,Kleinert2006-book} over all paths passing through the initial and final points,
    \begin{align}
    K(x_f,x_i,T) = \int_{x(0)=x_i}^{x(T)=x_f} \scD x~ \exp \frac{iS[x]}{\hbar} .
    \label{KFPI}
    \end{align}
Here, $S[x]=\int_0^T dt~ L(x,\dot x)$ is the action, 
$L(x,\dot x)=T(\dot x) - V(x)$ is the Lagrangian,
$T(\dot x)=M\dot{x}^2/2$ is the kinetic energy,
and $V(x)$ is the potential energy.  
Equation~\eqref{KFPI} has a simple and appealing interpretation, emphasized by Feynman himself.\cite{FeynmanLec2005-book,Goodman81v49,Schulman2005-book}
In this interpretation, the particle takes \emph{all possible paths} from the initial point to the final point.  Each path is associated with a complex phase factor $e^{iS/\hbar}$, where $S$ is the action accumulated along that path.  

\myheading{Semiclassical approximation.}
According to the correspondence principle, the predictions of quantum mechanics should reduce to those of classical mechanics when $x_i$, $x_f$, and $T$ are so large that $\hbar$ is negligible.  The \QB{} should reduce to the standard problem of a table tennis ball bouncing elastically and vertically on a table.  Indeed, one can develop the semiclassical approximation (SCA), in which the propagator is a sum over all classical paths $X_\alpha(t)$:
    \begin{align}
    K_\text{SCA} (x_f,x_i,T) 
    &= 
    \sqrt{\tfrac{i}{2\pi\hbar}}
    \sum_{\alpha}
    \sqrt{ \abs{ D_\alpha }}
    \exp i \left( 
        \frac{S_\alpha}{\hbar} - \frac{\pi m_\alpha}{2} \right).
    \label{KSemiclassical}
    \end{align}
For each classical path,
$S_\alpha = S[X_\alpha]$ is the action,
$D_\alpha 
= \det \frac{\dd^2 S_\alpha}{\dd x_i \dd x_f} 
\equiv  \frac{\dd^2 S_\alpha}{\dd x_i \dd x_f}$
is the van Vleck determinant,
and $m_\alpha$ is the Morse index.
Equations~\eqref{KFPI} and \eqref{KSemiclassical} are well known in the literature.  The reader may consult Schulman's book\cite{Schulman2005-book} for rigorous derivations. 
In this article we concentrate on applying these equations to the bouncers (i.e., \QB{} and \SB{}) and visualizing the results and procedure. 

\subsection{Classical paths}  \label{secClassicalPaths}
A \emph{stationary path} is defined as a path such that the action is stationary to first order with respect to small variations of the path.  Stationary paths obey the Euler-Lagrange equations $\frac{\dd L}{\dd x} = \frac{d}{dt} \frac{\dd L}{\dd \dot x}$, which reduce to the Newtonian equation of motion, $M\ddot{x} = -\frac{dV}{dx}$.  Thus, stationary paths are classical Newtonian trajectories satisfying the boundary conditions, and we will often refer to them as \emph{classical paths}.

\begin{figure}[htbp]
    \includegraphics[width=\textwidth]{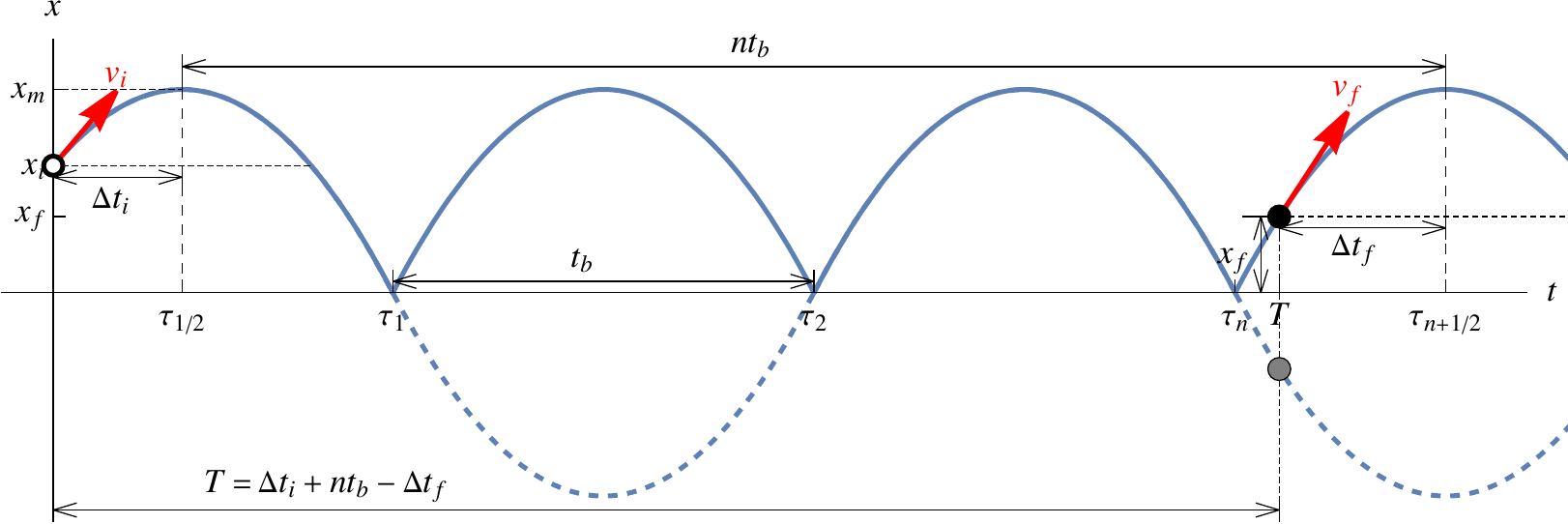}
    \caption{
    \label{Bouncer1Schematic}
    \textbf{Multiple-bounce path.}
    The solid curve shows a path of the \QB{} with $n=3$ bounces
    starting at $(0,x_i)$ and passing through $(T,x_f)$.
    The time between bounces is $t_b$.
    Bounce times are $\tau_1,\tau_2,\dotsc,\tau_n$
    and zenith times are $\tau_{1/2},\tau_{3/2},\dotsc,\tau_{n+1/2}$.
    The dashed curve shows a path of the \SB{}
        obtained by reflecting even-numbered segments of the original path.
    In this example $v_i$ and $v_f$ are positive,
        but in general they may take any real values.
    }
	\end{figure}

The \QB{} involves a particle of mass $M$ moving in a potential
$V(x) = Mgx + \infty \Theta(-x)$, where $\Theta$ is the unit step function.  While the particle is above the ground ($x>0$), it behaves as a freely falling particle with $\ddot{x} = -g$.  Whenever the particle meets the ground ($x=0$), it bounces elastically, so the velocity changes sign.  Solving these equations shows that a classical path with $n$ bounces ($n=0,1,2,\dotsc$) consists of $(n+1)$ parabolic segments.  See Fig.~\ref{Bouncer1Schematic}.

\myheading{Classical paths with $n$ bounces.}
From energy conservation, or from the kinematic equations of free fall, the maximum speed of the particle $v_m$ (attained at ground level) is related to the initial velocity $v_i$ and final velocity $v_f$ by
    \begin{align}
    v_m^2 = v_i^2 + 2gx_i = v_f^2 + 2gx_f.
    \label{EnergyConservation}
    \end{align}
The time interval between two consecutive bounces is $t_b=2v_m/g$.  The gravitational force $-Mg$ acting over time $T$ produces impulse $-MgT$, and each elastic collision with the ground produces impulse $+2Mv_m$.  Considering the total impulse leads to
    \begin{align}
    v_f = v_i - gT + 2n v_m.
    \label{Impulse}
    \end{align}
Alternatively, from Fig.~\ref{Bouncer1Schematic} we see that the time between initial and final points is
$T
= \Delta t_i + n t_b - \Delta t_f 
= \frac{v_i}{g} + n \frac{2v_m}{g} - \frac{v_f}{g}$,
 confirming Eq.~\eqref{Impulse}.

Using Eq.~\eqref{EnergyConservation}, we may eliminate $v_f$ and $v_m$ from Eq.~\eqref{Impulse} to obtain an equation relating $v_i$ to $x_i$, $x_f$, and $T$.  To simplify the algebra, we use the dimensionless variables 
$a=2x_i/gT^2$, $b=2x_f/gT^2$, $\tilvi = v_i/gT$:
    \begin{align}
    4n(\tilvi-1) \sqrt{\tilvi^2+a} + 4n^2\tilvi^2 - 2\tilvi
    + (4n^2-1)a+b+1 &= 0.
    \label{QuarticInDisguise}
    \end{align}
This leads to a polynomial equation for $\tilvi$:
    \begin{align}
    \scP(\tilvi) \equiv
    16n^2(n^2-1) \tilvi^4
    + 16n^2 \tilvi^3
    + [32n^4a + 8n^2(b-1-3a) + 4] \tilvi^2
    + 4(4n^2a+a-b-1) \tilvi
        \nonumber\\{}
    + [16n^4a^2 + 8n^2a(b-1-a) + (1+b-a)^2] = 0.
    \label{Quartic}
    \end{align}
Under different conditions, we need to solve either a quadratic, cubic, or quartic equation.  This means that a closed-form solution exists in terms of radicals (Ref.~\onlinecite{NIST-DLMF} Sec.~1.11), although for numerical computation, one should use library functions that handle special cases and minimize roundoff error.\cite{NumericalRecipes}

    
Once we have found $v_i$, we can calculate the 
maximum speed $v_m=\sqrt{v_i^2+2gx_i}$, 
time between bounces $t_b=2v_m/g$, and 
maximum height $x_m={v_m}^2/2g$.
From Fig.~\ref{Bouncer1Schematic}, the time of the $k$th bounce is 
$\tau_k = \Delta t_i + (k-\half)t_b =[ v_i + (2k-1)v_m]/g$
where $k=1,2,3,\dotsc,n$.
The time of the $k$th zenith (highest point) is 
$\tau_{k+1/2} = (v_i + 2kv_m)/g$.
Note that $\tau_{1/2}$ and $\tau_{n+1/2}$ may or may not lie within the interval $[0,T]$. 
The index of the zenith closest to time $t$ is $k=\floor{\frac{gt-v_i}{2v_m} + \frac{1}{2}}$, where $\lfloor\dots\rfloor$ is the floor function.  The path is $X(t) = x_m - \frac{g}{2} (t - \tau_{k+1/2} )^2$.  In terms of the original variables,
    \begin{align}
    X(t) = x_i + \frac{{v_i}^2}{2g} 
    - \frac{g}{2} \left(
        t - \frac{v_i}{g} 
        - \frac{2\sqrt{v_i^2+2gx_i}}{g}
            \floor*{ \frac{gt-v_i}{2\sqrt{v_i^2+2gx_i}} + \frac{1}{2} }
        \right)^2
        .
    \label{Xt}
    \end{align}

\myheading{Phase diagram.}
The discriminant of the polynomial $\scP(\tilvi)$ in Eq.~\eqref{Quartic} is
    \begin{align}
    \scD_n(a,b)
    &=  4096 c^2 \big[ b + (a+1)(c-1) \big]^2 \times
    \big[
    1+(2-2 c) (a+b)
            \nonumber\\&
    +\left(1-8 c+c^2\right) (a+b)^2+\left(-4+20 c+2 c^2\right) a b
            \nonumber\\&
    +\left(-10 c+2 c^2\right) (a+b)^3+\left(40 c-2 c^2-2 c^3\right) a b (a+b)
            \nonumber\\&
    +\left(-64
   c+48 c^2-12 c^3+c^4\right) a^2 b^2+\left(32 c-16 c^2+2 c^3\right) a b (a+b)^2+\left(-4 c+c^2\right) (a+b)^4
    \big]
    \label{discriminant}
    \end{align}
where $c=4n^2$.  The last factor is symmetric in $a$ and $b$.  The sign of the discriminant gives information about the number of solutions for $v_i$, that is, the number of classical paths with $n$ bounces:

\begin{itemize}[noitemsep]
\item
If $n=0$, $\scP(\tilvi)$ is quadratic, and $\scD_0=0$, indicating that the quadratic equation has only one unique root, which is $\tilvi=(1+b-a)/2$.
\item
If $n=1$, $\scP(\tilvi)$ is cubic.  If $\scD_1<0$ there is only 1 real root, and hence 1 path.  If $\scD_1>0$ there are 3 real roots (3 paths).  
\item
If $n=2,3,4,\dotsc$, $\scP(\tilvi)$ is generally quartic.  If $\scD_n>0$ there are either 0 or 4 real roots.  If $\scD_n<0$ there are 2 real roots.
\end{itemize}

\begin{figure}[htbp]
    \includegraphics[width=\textwidth]{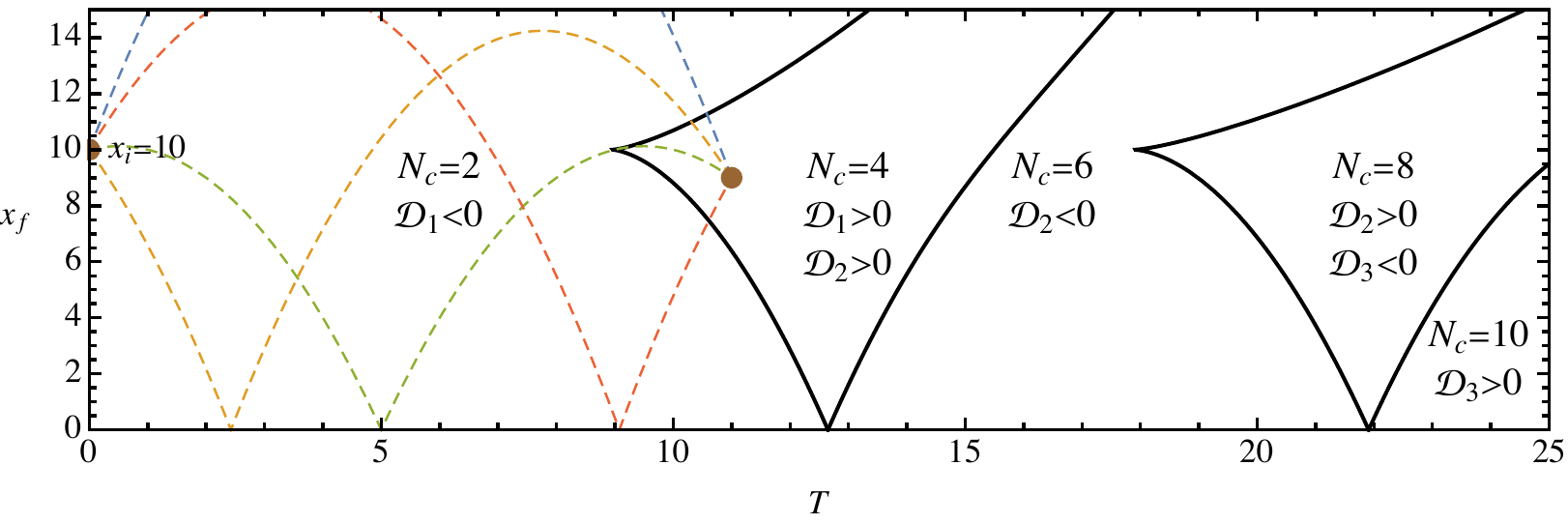}
    \caption{
    \label{Bouncer1PhaseDiagram}
    \textbf{Phase diagram of the classical one-sided bouncer.}
    This diagram shows the number of classical paths $N_c$ as a function of final position $x_f$ and final time $T$, for fixed $x_i=10$.  
    For example, the dot at $(T,x_f)=(11,9)$ lies within the $N_c=4$ region, so there are four classical paths, shown as dashed curves.  One of these paths has $n=0$ bounce, and the remaining three paths have $n=1$ bounce.
    }
	\end{figure}

For a fixed number of bounces, $n$, there are at most four paths.
As $T$ increases, it becomes possible to find classical paths with larger $n$.  Therefore the total number of classical paths $N_c$ increases.
We may construct a phase diagram in $(x_i,x_f,T)$ space by plotting the surfaces where $\scD_n(2x_i/gT^2,2x_f/gT^2)=0$.  Figure~\ref{Bouncer1PhaseDiagram} shows the phase diagram in the $(x_f,T)$ plane for a particular value of $x_i$.  

The phase boundaries are complicated quartic curves, but there are simple expressions for the positions of the ``kinks.''  
If $\scD_n(a,0)=0$, then $a=1/4n(n-1)$.
If $\scD_n(a,a)=0$, then $a=1/4(n^2-1)$.
By examining the shape of the zero contours of $\scD_n(a,b)$, one can derive a simple upper bound on the number of bounces, $n \leq \sqrt{1 + 1/(2a+2b)}$.  Thus, for given $a$ and $b$, one only needs to solve the quartic equation for
    \begin{align}
    n = 0,1,2,\dotsc,\ceil{\sqrt{1 + 1/(2a+2b)}}
    .
    \label{BoundOnBounces}
    \end{align}
 
Equations~\eqref{Quartic}, \eqref{Xt}, and \eqref{BoundOnBounces} represent the solution to the \emph{path enumeration problem} --- the problem of finding all classical paths $X(t)$ beginning at $X(0)=x_i$ and ending at $X(T)=x_f$.

\subsection{Classical action}
\myheading{Zero-bounce path.}
First consider a particle with uniform acceleration $\ddot{X}=-g$, initial position $X(0)=x_i$, and final position $X(T)=x_f$.  Solving the differential equation gives
    $
    X(t)
    =\frac{T-t}{T} x_i + \frac{t}{T} x_f + \frac{gt(T-t)}{2}
    $.
The action, $S=\int_0^T dt~ L(X,\dot X)$, works out to be
	$
    S
    =\frac{M}{2T} (x_f - x_i)^2 
    - \frac{MgT}{2} (x_f + x_i)
    - \frac{M g^2 T^3}{24}
    $.
Using Eqs.~\eqref{EnergyConservation} and \eqref{Impulse}, we may rewrite the action in the useful form
	\begin{align}
    S(v_i,v_f,v_m)
    &=\frac{M}{6g} \left[
        2({v_i}^3 - {v_f}^3) - 3{v_m}^2 (v_i - v_f)
        \right].
    \label{SFromVelocities}
    \end{align}

\myheading{Multi-bounce path.}
For a path with $n$ bounces, the action associated with each bounce is negligible.  (This can be proven by replacing the infinite potential barrier by a steep ramp, calculating the action during the time that the particle is on the ramp, and taking the limit as the slope of the ramp goes to infinity.)
Thus the total action is simply the sum of contributions from each parabolic segment.  Each segment has the same value of $v_m$.  The first segment has initial velocity $v_i$ and final velocity $-v_m$.  Each intermediate segment has initial velocity $+v_m$ and final velocity $-v_m$.  The final segment has initial velocity $+v_m$ and final velocity $v_f$.  Applying Eq.~\eqref{SFromVelocities} to each segment gives the total action as
$S(v_i, -v_m, v_m) + (n-1) S(v_m, -v_m, v_m) + S(v_m, v_f, v_m)$.
After some algebra, this reduces to
	\begin{align}
	S(n,v_i,v_f,v_m)
    &=\frac{M}{6g} \left[
        2({v_i}^3 - {v_f}^3) - 3{v_m}^2 (v_i - v_f) - 2n {v_m}^3
        \right].
    \label{SFromVelocitiesFull}
    \end{align}

\newcommand{\defeq}{\overset{\mathrm{def}}{=}}

\subsection{Van Vleck determinant}
For motion in one dimension, the van Vleck ``determinant'' $D = \frac{\dd^2 S}{\dd x_i ~ \dd x_f}$ is simply the mixed second derivative of the action $S(x_i,x_f,T)$.  We compute $D$ explicitly below as follows.  (In this section, $n$ is treated as a constant.)

We first perform implicit differentiation on Eq.~\eqref{QuarticInDisguise} to obtain a relation between $d\tilvi$, $da$, and $db$.  This leads to
	\begin{align}
	\frac{\dd \tilvi}{\dd a}
    &=
    \frac{2 n (1-u)-\left(4 n^2-1\right) \sqrt{a+u^2}}{2 \sqrt{a+u^2} \left(4 n^2 u-1\right)+4 n \left(a+2 u^2-u\right)}
    \label{duda}
        ,\\
    \frac{\dd \tilvi}{\dd b}
    &=
    \frac{ - \sqrt{a+u^2}}{2 \sqrt{a+u^2} \left(4 n^2 u-1\right)+4 n \left(a+2 u^2-u\right)}
        .   
    \label{dudb}
    \end{align}
Substituting Eqs.~\eqref{EnergyConservation} and \eqref{Impulse} into Eq.~\eqref{SFromVelocitiesFull} gives the action as a function of $x_i$ and $v_i$ only.  In terms of dimensionless variables,
	\begin{align}
	S(a,\tilvi)
    &=
    2-3 a-6 u+3 u^2+24 n^2 \left(a+u^2\right) (1-u)
        \nonumber\\&~~~{}
    +n \left[ -12+4 a+24 u-8 u^2-16 n^2 \left(a+u^2\right)\right] \sqrt{a+u^2}
        .   
    \label{Sau}
    \end{align}
Differentiate this twice using the chain rule:
	\begin{align}
	\frac{\dd S(a,b)}{\dd a}
    &= 
        \frac{\dd S(a,\tilvi)}{\dd a}
    +   \frac{\dd \tilvi}{\dd a} \frac{\dd S(a,\tilvi)}{\dd \tilvi}
    \defeq
        F(a,\tilvi)
        ,\nonumber\\
    \frac{\dd^2 S(a,b)}{\dd a ~ \dd b}
    &= 
        \frac{\dd F(a,\tilvi)}{\dd b}
    =
         \frac{\dd \tilvi}{\dd b}   \frac{\dd F(a,\tilvi)}{\dd \tilvi}.
    \end{align}
With the help of Mathematica, and after much manipulation, we obtain a remarkably simple expression,
    \begin{align}
	D
    &= 
        \frac{\dd^2 S}{\dd x_i ~ \dd x_f}
    =   
        \frac{4}{g^2 T^2}   \frac{\dd^2 S}{\dd a~ \dd b}
    =
        \frac{ M {v_m}^2 }{ T  v_i v_f + 2(x_i v_f - x_f v_i)}
        .
    \label{VVD}
    \end{align}
Note that for a given set of parameters $(x_i,x_f,T)$, there may be more than one path, $X_\alpha(t)$, passing through the end points;  for each path there is a corresponding value of the van Vleck determinant, $D_\alpha$. 

\begin{figure}[htbp]
    \includegraphics[width=\textwidth]{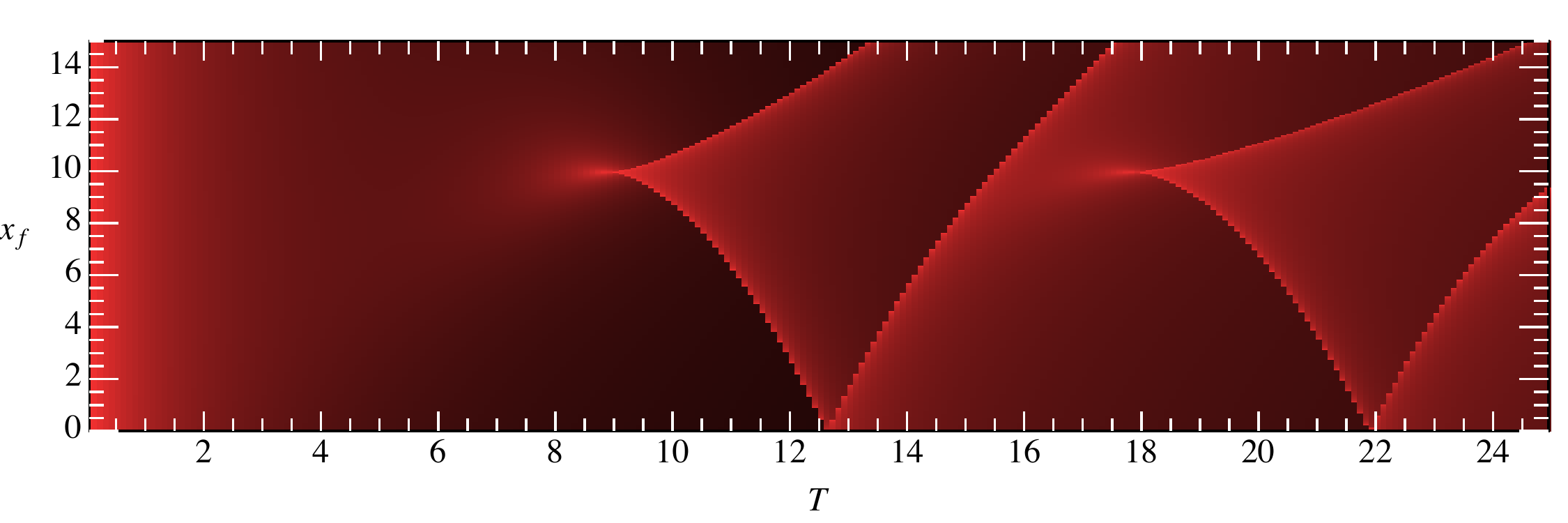}
    \caption{
    \label{Bouncer1VVD}
    Heatmap of $K_\text{VVD} (x_f,x_i,T)$ for $x_i=10$. 
    This quantity is the sum of the amplitudes of contributions to the propagator from each path, omitting phase factors.
    }
	\end{figure}

As described in the literature, $\abs{D_\alpha}$ describes the density of paths in the vicinity of a chosen path.  It determines the amplitude of the contribution of each path to the propagator.  Let
    $
    K_\text{VVD}
    =
    \sqrt{\tfrac{1}{2\pi\hbar}}
    \sum_{\alpha}
    \sqrt{ \abs{ D_\alpha }}
    $
be the sum of the amplitudes, omitting the phase factors that appear in Eq.~\eqref{KSemiclassical}.  This is an envelope function such that $\abs{K} \leq K_\text{VVD}$.
Figure \ref{Bouncer1VVD} shows a heatmap of $K_\text{VVD}$
in the $(T,x_f)$ plane.  This quantity is large near the phase boundaries in Fig.~\ref{Bouncer1PhaseDiagram}. 

\subsection{Morse index} \label{secMorseIndex1}
The last ingredient in Eq.~\eqref{KSemiclassical} is the Morse index $m_\alpha$, defined here as the number of foci along stationary path $X_\alpha(t)$.  
Sometimes the Morse index is also known as the Maslov-Morse index; see e.g., Ref.~\onlinecite{Kleinert2006-book}.
The Morse index for the \QB{} is easiest to derive from the Morse index for the \SB{}, which is discussed in detail in Sec.~\ref{secMorseIndex2}.
The reader may wish to read Sec.~\ref{secMapping} and Sec.~\ref{secMorseIndex2} before returning to this point.

From Eq.~\eqref{KSemiclassical} we know that each focus contributes a phase difference of $-\pi/2$ to the propagator.  When mapping paths of the \SB{} back to the \QB{}, each bounce is equivalent to two additional foci.  
In the language of the WKB approximation, reflection from a classically forbidden region corresponds to a single turning point, introducing a phase change of $-\pi/2$, whereas reflection from a hard wall (infinite potential barrier) \cite{Schulman2005-book} introduces a phase change of $\pi$.
In Sec.~\ref{secGoodmanSubtraction} we provide yet another way to understand this result.  
Ultimately, the formula for the Morse index for a path of the \QB{}, Eq.~\eqref{MorseIndex1}, is given by taking Eq.~\eqref{MorseIndex2} and adding twice the number of bounces ($n$):
    \begin{align}
    v_m &= \sqrt{{v_i}^2 + 2gx_i}  ,\\
    n &= \left\lfloor  \frac{gT-v_i}{2v_m} + \frac{1}{2} \right\rfloor , \\
    j &= \left\lfloor  \frac{gT-v_i}{2v_m} \right\rfloor  + \Theta(v_i)  ,\\
    k &= j + \Theta \left( -\sqrt{gx_i} - \sqrt{2j(j+1)} v_i  \right)  ,\\
    m &= j + \Theta \left( T -
        \frac{2k}{g} \frac{{v_i}^2 + {v_m}^2 + 2k v_i v_m}{v_m + 2k v_i}  
        \right)
        + 2 n
    .
    \label{MorseIndex1}
    \end{align}

We reiterate at this point that a rigorous justification of the Morse index calculation based on Eq.~\ref{MorseIndex1} for the \QB{} is similar to the justification of the Morse index calculation based on Eq.~\ref{MorseIndex2} for the \SB{}. More details
are given for the \SB{} since the discussion is easier compared to that for the \QB{}.

\subsection{Propagator}
We can now calculate the semiclassical propagator $K_\text{SCA} (x_f, x_i, T)$ as follows.
Loop over $n$ according to Eq.~\eqref{BoundOnBounces}.  Find allowed values of $v_{i\alpha}$ by solving Eq.~\eqref{Quartic}.  Find the corresponding values of $S_\alpha$, $D_\alpha$, and $m_\alpha$ using Eqs.~\eqref{SFromVelocitiesFull}, \eqref{VVD}, and \eqref{MorseIndex1}.  Finally, find $K$ as a sum over all classical paths using Eq.~\eqref{KSemiclassical}.

    \begin{figure}[htbp]
    \includegraphics[width=\textwidth]{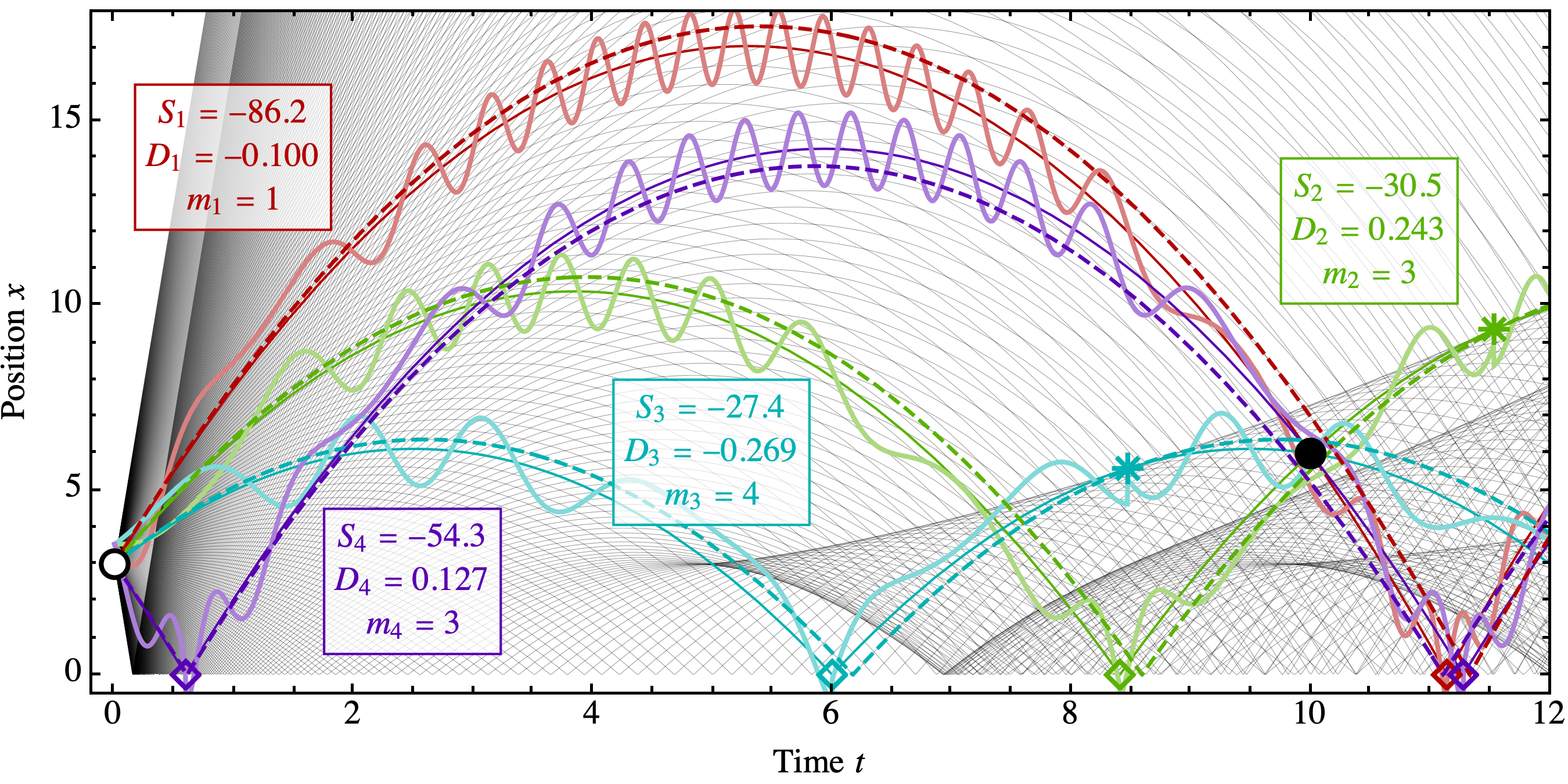}
    \caption{
    \label{Bouncer1Everything}
    \textbf{Visualizing the semiclassical approximation for the \QB{}.}
    This figure gives a quantitative and illuminating picture of the meaning of the action $S_\alpha$, the van Vleck determinant $D_\alpha$, and the Morse index $m_\alpha$ that occur in the semiclassical propagator $K_\text{SCA}$.
    Parameters are $x_i=3$, $x_f=6$, $T=10$, and $M=g=\hbar=1$.
    The total number of classical paths is $N_c=4$.
    There is one path with $n=0$ bounce, and there are three paths with $n=1$ bounce.
    }
	\end{figure}

\subsection{Visualizing the Semiclassical Approximation}
In most introductory courses on optics or modern physics, Young's double-slit experiment is explained in terms of a semiclassical path integral.  A photon (or electron) passes simultaneously through two slits.  It travels along two paths, accumulating a different phase along each path as it propagates through air (or through optical elements or electric fields inserted along the path).  At each point on the screen, constructive or destructive interference occurs depending on the \emph{phase difference}.  The visibility of the interference fringes depends upon the relative \emph{amplitudes} of the rays from the slits.

Figure \ref{Bouncer1Everything} explains the \QB{} in terms of these very same concepts.
Suppose a quantum particle (``bouncing ball'') is released at position $x_i=3$ at time $t=0$, and is subsequently detected at position $x_f=6$ at time $T=10$.
Classically, it must have taken one of four paths ($\alpha=1,2,3,4$).

\myheading{Phase.}
According to Eq.~\ref{KSemiclassical}, we see that a particle traveling along path $X_\alpha(t)$ picks up a time-dependent phase factor $\exp i \theta_\alpha(t)$,
where the \emph{phase angle} is
$\theta_\alpha(t) = \frac{S_\alpha(t)}{\hbar} - \frac{\pi m_\alpha(t)}{2}$.
The first contribution to $\theta$ comes from the \emph{classical action}, which describes propagation in time under the influence of the potential.  If $S_\alpha=Nh$ (where $h$ is Planck's constant), then $\theta_\alpha=2\pi N$, so the phase goes through $N$ complete cycles.  
The second contribution to $\theta$ is due to the \emph{Morse index}.  In Fig.~\ref{Bouncer1Everything}, dashed curves show paths resulting from slight perturbations of the initial velocity $v_i$.  The dashed curves and solid curves drift apart initially, but they converge again at foci, indicated by stars along each classical path.  The Morse index $m_\alpha$ is the number of foci along path $X_\alpha$ within the time interval $[0,T]$.  The initial point $(0,x_i)$ counts as one focus.  A bounce, indicated by a diamond, counts as \emph{two} foci.  For example, the blue curve (the third classical path $X_3$) has $m_3 = 1+2+1 = 4$.
In Fig.~\ref{Bouncer1Everything}, the phase is visualized as sinusoidal wiggles with discontinuous jumps at bounces and foci.


\myheading{Amplitude.}  
From Eq.~\ref{KSemiclassical}, the amplitude of the contribution from path $\alpha$ is $\sqrt{ \abs{ D_\alpha } / 2\pi\hbar}$ where $D_\alpha$ is the van Vleck determinant.  It may be shown that $D_\alpha(t) = -1/f_\alpha(t)$ where the path divergence function, $f_\alpha(t) = \dd X(t)/\dd v_i$, is the deviation between the solid curve and the dashed curve.  If path $\alpha$ has a focus near time $T$, then $f_\alpha(T)$ is small and $D_\alpha(T)$ is large.

    \begin{figure*}[htbp]
    \includegraphics[width=1\textwidth]{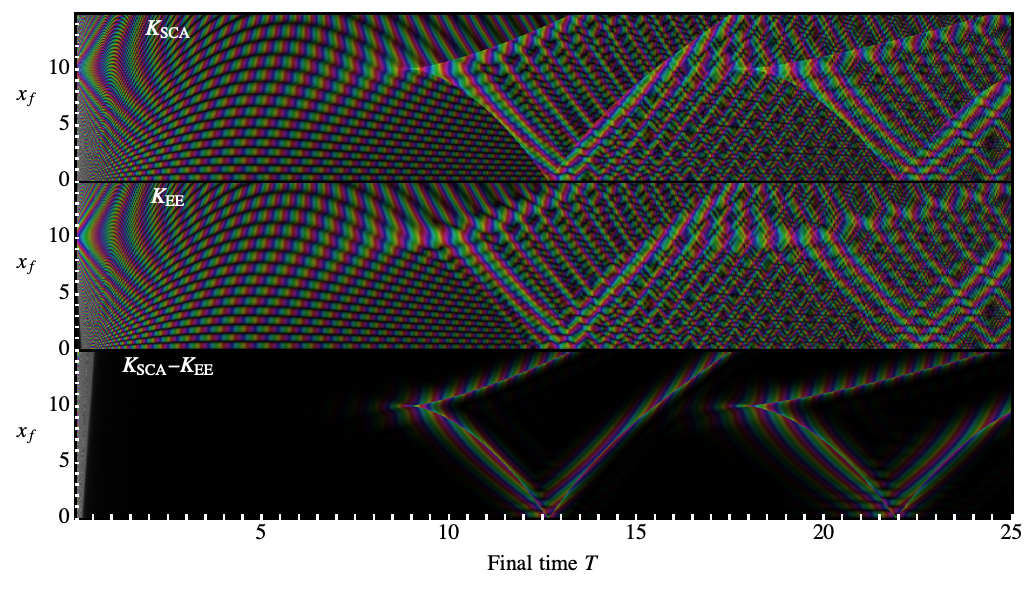}
    \caption{
    \label{Bouncer1Heatmaps}
    \textbf{Propagator $K(x_f,x_i,T)$ for \QB{} for $x_i=10$.}
    (Top) Propagator from \PI{} in semiclassical approximation, $K_\text{SCA}$, shown as heatmap with rainbow palette for complex magnitude and argument.
    (Center) Propagator calculated using EE truncated to $10^4$ Airy functions.
    (Bottom) Discrepancy $K_\text{SCA} - K_\text{EE}$.
    }	      
	\end{figure*}

    \begin{figure}[htbp]
    \includegraphics[width=1\textwidth]{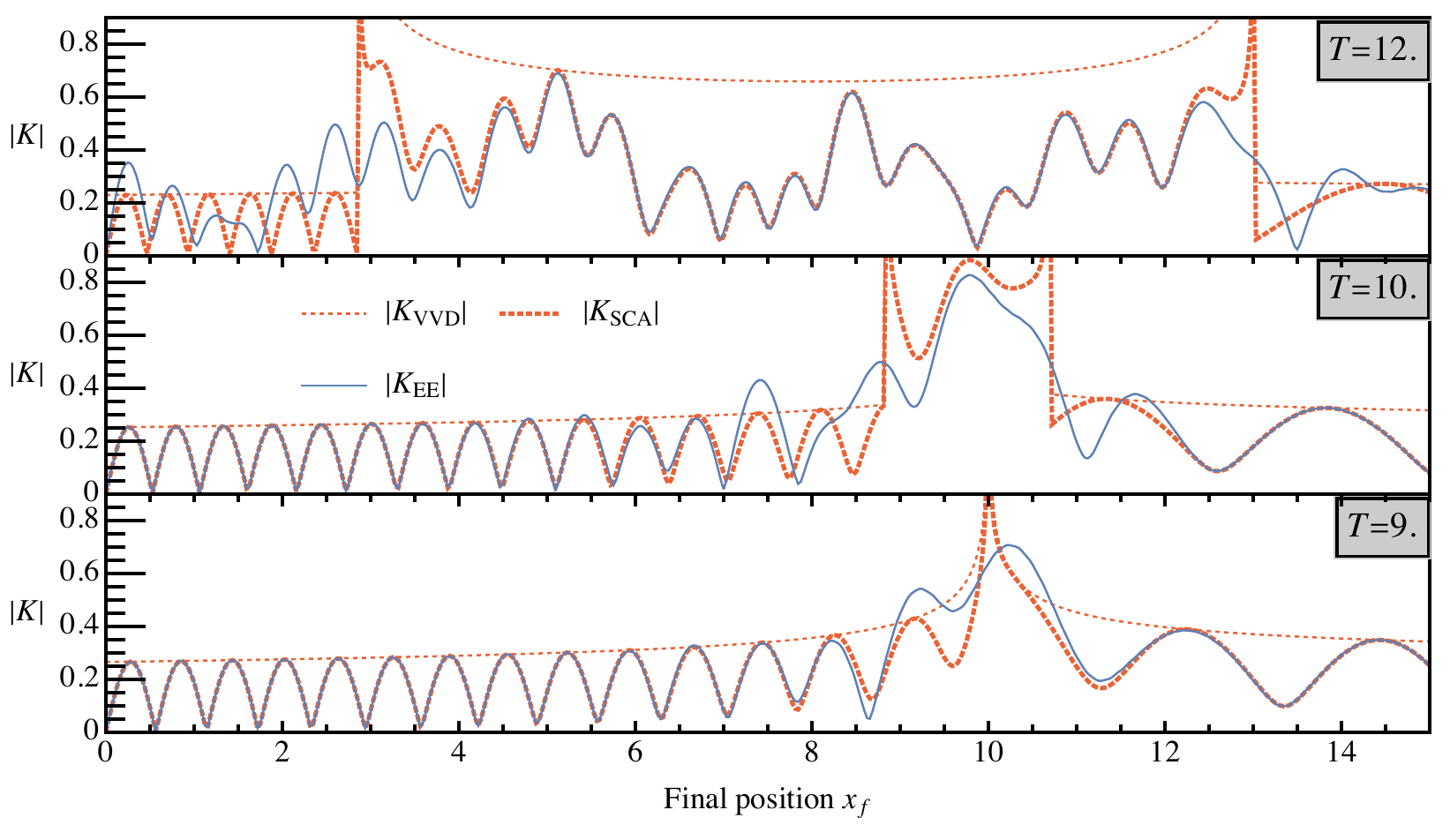}
    \caption{
    \label{Bouncer1Slices}
    Propagator $K(x_f,x_i,T)$ for \QB{} for $x_i=10$ at various times $T$.  
    $K_\text{SCA}$ is the propagator in the semiclassical approximation.
    $K_\text{VVD}$ is the envelope obtained by ignoring phases.
    $K_\text{EE}$ is the propagator from the EE method.
    }
	\end{figure}
	
\subsection{Comparison between \PI{}-SCA and EE}
Figures~\ref{Bouncer1Heatmaps} and \ref{Bouncer1Slices} compare the \QB{} propagator for $x_i=10$, calculated numerically using the \PI{}-SCA and EE methods.
The EE method converges fairly well for most values of $T$ and $x_f$ shown in the figure.  However, at small $T$ ($T<0.25$) and large $x_f$ ($x_f>10$), convergence is poor.  Even $10^4$ terms are insufficient.
In contrast, the \PI{} method is very accurate at small $T$.  When the time of flight is short, gravity has negligible effect, and the classical paths resemble free-particle paths.  In this limit the bouncer is equivalent to a particle near a hard wall.  The propagator is the difference between a ``direct'' contribution (where the particle goes directly from $x_i$ to $x_f$) and a ``single-bounce'' contribution (where the particle moves at high speed and bounces off the ground).  


The reader may be surprised that the propagator bears very little resemblance to the time evolution of the a Gaussian wave packet shown in Appendix~\ref{secWavePacketEvolution}.  Instead, the structure of the propagator is dictated by the phase boundaries in Fig.~\ref{Bouncer1PhaseDiagram}.  The nature of the propagator changes suddenly over a few wavelengths every time a pair of new classical paths begins to contribute.


\section{Symmetric Bouncer}  \label{secBouncer2}

\subsection{Eigenfunction Expansion Method} \label{secEE2}
We now turn to the symmetric bouncer (\SB{}), described by the Hamiltonian $\hat{H} = \hat{p}^2/2M + V_s(\hat{x})$ with the potential energy function shown in Fig.~\ref{VSymmetrized},
    \begin{align}
    V_s(x) &= M g \abs{x}.
	\label{eqVs}
	\end{align}
Solving the Schr\"odinger equation, 
$-\frac{\hbar^2}{2M} \frac{d^2\psi}{dx^2} + M g \abs{x} \psi = E \psi$,
we find that the even- and odd-parity eigenfunctions and the corresponding energies are
    \begin{align}
    E_m^{(+)} &= -\mu_m \frac{E_0}{\gamma}
        ,\\
	E_m^{(-)} &= -\lambda_m \frac{E_0}{\gamma}
        ,\\
    \varphi_m^{(+)} 
    &= \sqrt{\frac{\gamma}{2x_0(-\mu_m)} } 
        \frac{\Ai(\frac{\gamma\abs{x}}{x_0} + \mu_m)}{\Ai(\mu_m)  }    
        ,\\
	\varphi_m^{(-)} 
    &= \sqrt{\frac{\gamma}{2x_0} } 
        \frac{\Ai(\frac{\gamma\abs{x}}{x_0} + \lambda_m)}{ \Ai'(\lambda_m)  } 
        \sgn x
        .
	\end{align}
As before, $\gamma=2^{1/3}$ and $\lambda_m$ is the $m$th zero of the Airy function.  Here, $\mu_m$ is the $m$th zero of the derivative of the Airy function, such that $\Ai'(\mu_m)=0$.
Henceforth, we will refer to all eigenenergies and eigenfunctions (both even and odd) collectively as $E_n$ and $\varphi_n$.
See Fig.~\ref{Bouncer2Eigen}.
As before, the propagator can be written as
    \begin{align}
    K_s (x_f, x_i, T)
    &= 
        \sum_{n=1}^\infty
        \varphi_n^{(\pm)} (x_f) e^{-i E_n^{(\pm)} T/\hbar} \varphi_n^{(\pm)*} (x_i).
    \label{eqKEE2}
	\end{align}

\subsection{Feynman Path Integral: Mapping between \SB{} and \QB}  \label{secMapping}
Suppose $X(t)$ is a classical path of the \QB{} with $n$ bounces, initial position $x_i$, final position $x_f$, and initial velocity $v_i$.  This path $X(t)$ corresponds to two classical paths of the \SB{}, $X^s_\pm (t)$, with parameters
$n^s = n$, 
$x_i^s = \pm x_i$,
$x_f^s = \pm x_f (-1)^n$, and
$v_i^s = \pm v_i$.
The dashed curve in Fig.~\ref{Bouncer1Schematic} shows $X^s_{+} (t)$. 

In the \SB{}, the paths do not reflect from the boundary $x=0$, but pass straight through it.  Nevertheless, we will continue to refer to zero-crossings of the path as ``bounces''.  After all, the potential $V_s(x)$ can describe a marble rolling in a V-shaped track.  When the marble experience a sudden change of force at $x=0$, this appears similar to a bounce.

Suppose $X_s(t)$ is a classical path of the \SB{} with $n^s$ ``bounces,'' initial position $x^s_i$, final position $x^s_f$, and initial velocity $v^s_i$.  Note that the path changes sign after every bounce.  Therefore, if $n_s$ is even, the initial and final positions must have the same sign, $x_i x_f>0$.  Conversely, if $n_s$ is odd, then $x_i x_f < 0$.  Assuming this criterion is satisfied, we can invert the mapping to obtain the parameters of the \QB{} as
    \begin{align}
    n &= n^s                                &\qquad  
    x_i &= \abs{x_i^s}                       \nonumber\\
    x_f &= (-1)^n  x_f^s  \sgn x_i^s        &\qquad
    v_i &= v_i^s \sgn x_i^s.
    \label{BackwardMapping}
    \end{align}
It is easy to see that the action for a path of the \SB{} is the same as the action for the corresponding path of the \QB{} obtained from the mapping above.  Thus we may write $S^s_\alpha = S_\alpha$.  $D_\alpha$ may experience a sign change due to the mapping, depending on whether $x_i$ and/or $x_f$ are reflected; however, $\abs{D_\alpha}$ is not affected.

\subsection{Morse index}  \label{secMorseIndex2}
Here we derive a formula for the Morse index (the number of foci along a given classical path) for the \SB.

\myheading{Path divergence function.}
First consider a general action
    \begin{align}
    S[x]
    &=\int_0^T dt~ \left[ \frac{M}{2} \dot{x}^2 - V(x) \right]
    .
	\end{align}
Let $x(t) = X(t) + \xi(t)$ where $X(t)$ is a given classical path and $\xi(t)$ is a small variation that vanishes at the end points.  Expanding $S[x]$ to second order in $\xi$, performing integration by parts on the $\dot{X} \dot{\xi}$ cross-term, using the boundary conditions $\xi(0)=\xi(T)=0$, and invoking the Euler-Lagrange equation $M\ddot{X} = -V'(X)$, we find that $S[x] = S[X] + S[\xi]$ where
    $
    S[\xi]
    =\int_0^T dt~ \left[ \frac{M}{2} \dot{\xi}^2 
        - W(\xi) \right].
    $
In other words, the variation is governed by the \emph{residual potential}
    $
    W(\xi) = V(X+\xi) - V(x) - V'(X)\xi
    .
    $
Performing a Taylor expansion and keeping terms only up to $O(\xi^2)$, we obtain a harmonic action
    $
    S[\xi]
    =\int_0^T dt~ \left[ \frac{M}{2} \dot{\xi}^2 
    - \frac{\kappa(t)}{2} \xi^2 \right]
	$
where 
    $
    \kappa(t) = V''(X(t))
    $
is the local potential curvature at position $X(t)$ along the classical path at time $t$.  The Euler-Lagrange equation for $\xi$ is  
    $
    M \ddot \xi + \kappa(t) \xi = 0
    $.
Let the \emph{path divergence function}
$f(t) 
= \frac{\delta x(t)}{\delta v_i}
= \frac{\xi(t)}{\delta v_i}
$ 
be the deviation at time $t$ divided by the perturbation to the initial velocity $\delta v_i$.  In other words, $f(t)$ represents the \emph{sensitivity} of the path to the initial velocity.  We see that $f(t)$ satisfies the ODE
    \begin{align}
    M \ddot f + \kappa(t) f &= 0
    \end{align}
with boundary conditions $f(0)=0$ and $f'(0)=1$.

\begin{figure}[htbp]
    \includegraphics[width=\textwidth]{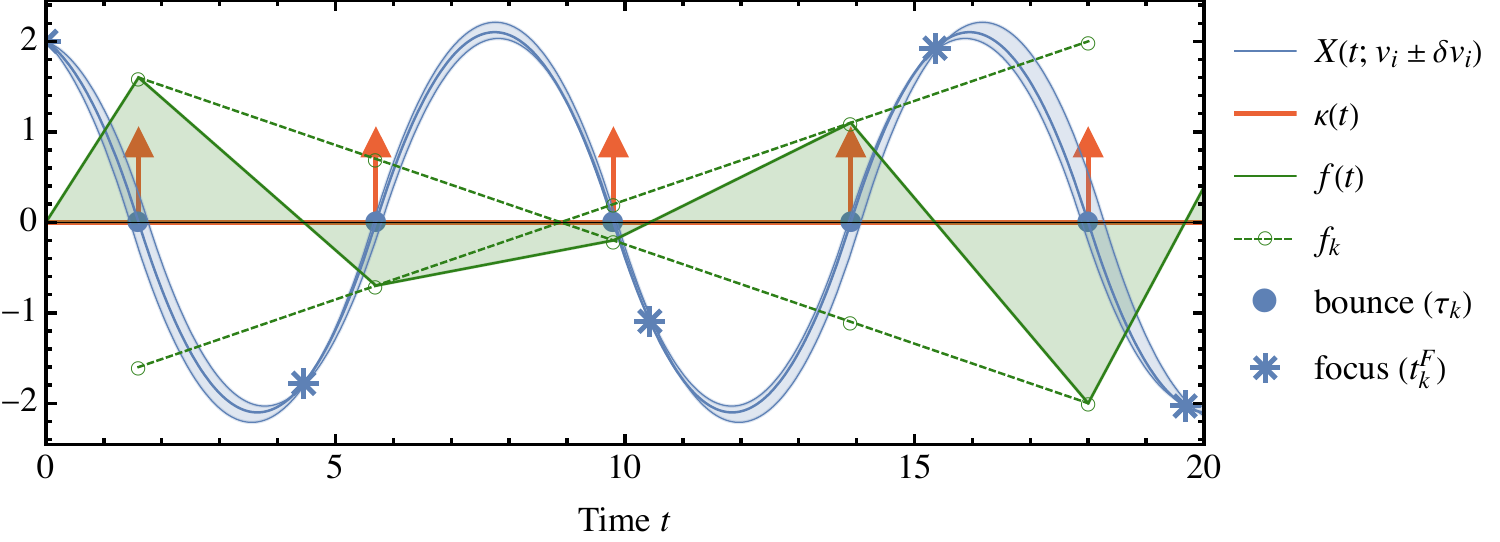}
    \caption{
    \label{PathDivergenceFunction}
    \textbf{Locating foci.}
    The blue solid curve shows the classical path $X(t)$ for $x_i=2$ and $v_i=-0.45$.  The blue band results from perturbing the initial velocity by $\pm\delta v_i = \pm 0.2$.  The bounce times are defined by $X(\tau_k)=0$.  The potential curvature function $\kappa(t)$ is a sum of Dirac delta functions.  The path divergence function $f(t)$ results from solving the 1D oscillator equation with a time-dependent spring constant $\kappa(t)$.  The green zigzag lines show $f(t)$.
    Whenever $f(t)=0$, the original path and the perturbed path converge at a focus, provided that $\delta v_i$ is small.
    $f(t)$ is a linear piecewise function passing through points $(\tau_k,f_k)$.  For these parameters, the envelope of $f_k$ passes through zero at $k_0 \approx 2.78$.  Therefore $f_k$ experiences a sign change and $f(t)=0$ has a root (focus) between every pair of bounces, \emph{except} between $\tau_2$ and $\tau_3$.
    }
	\end{figure}

For the \SB{}, we have $V(x)=Mg\abs{x}$.  Differentiating this twice gives a Dirac delta function, $V''(x) = 2Mg~ \delta(x)$.  Recall that the time of the $k$th bounce is $\tau_k = [ v_i + (2k-1)v_m]/g$, where $k=1,2,3,\dotsc,n$.  The velocity at each bounce is $\abs{dX/dt} = v_m$.  Using the rules for Dirac delta functions, we obtain
    \begin{align}
    \kappa(t) = V''(X(t)) 
    = 2Mg \sum_{k=-\infty}^\infty 
        \frac{\delta(t-\tau_k)}{ \abs{dX/dt} }
    = \frac{2Mg}{v_m} \sum_{k=-\infty}^\infty \delta(t-\tau_k).
    \end{align}
Since $\kappa(t)=0$ between bounces, $f(t)$ is a linear function between bounces.  Let $f(t)$ be a piecewise linear function such that $f(\tau_k) = f_k$ for $k=1,2,3,\dotsc$.  By considering the strength of the delta function, we see that $f'$ increases by $\frac{2g}{v_m} f(\tau_k)$ due to the $k$th bounce.  This allows us to develop a three-term recursion relation for $f_k$.  Solving this recursion relation analytically gives
    \begin{align}
    f_k = \frac{(-1)^{k+1}}{g} \left[ v_m + (2k-1) v_i \right].
    \label{fk}
    \end{align}
If desired, one can plot the piecewise linear function $f(t)$, as in Fig.~\ref{PathDivergenceFunction}.

Even though our derivations of the path divergence function $f(t) = \frac{\dd x(t)}{\dd v_i}$ and the van Vleck determinant $D = \frac{\dd^2 S}{\dd x_i ~ \dd x_f}$ seem completely different, these two quantities satisfy\cite{Schulman2005-book} the simple relation $D = -1/f$.  Thus, Eq.~\eqref{VVD} implies that
    \begin{align}
    f(t) &= 
        -\frac{t v_i v_f + 2(x_i v_f - x_f v_i) }{M v_m^2}.
    \label{ftExplicit}
    \end{align}
We find that this indeed agrees with Eq.~\eqref{fk}.
    
\myheading{Locating foci for \SB{}.}
Note that $v_m + (2k-1) v_i$ is a monotonic function of $k$ with only one zero, which is $k_0 = (1 - v_m/v_i)/2$.  Therefore, $f_k$ alternates in sign between every two integer values $k$ and $k+1$, \emph{unless} $k < k_0 < k+1$.  If $f_k$ and $f_{k+1}$ have opposite signs, then $f(t)=0$ has a root in the interval $(\tau_k,\tau_{k+1})$, which can be found by linear interpolation.  Ultimately, we find that $f(t)=0$ has roots at `focal' times $t^F_k$ such that
    \begin{align}
    t^F_k &= \left( \frac{2k}{g} \right) \frac{{v_i}^2 + {v_m}^2 + 2k v_i v_m}{v_m + 2k v_i}
        ,\quad 
    k=1,2,3,\dotsc
        ,\quad
    k \neq \floor{(1 - v_m/v_i)/2}
    .
    \end{align}
The path divergence function passes through zero at times $t^F_k$.  This means that paths with different initial velocities will converge and intersect at times $t^F_k$ (according to this first-order perturbative analysis).  In other words, \emph{foci} occur at these times.  To avoid having to exclude a certain value of $k$, we prefer to write
    \begin{align}
    t^F_k &= 
        \begin{cases}
        \frac{2(k+1)}{g} \frac{{v_i}^2 + {v_m}^2 + 2(k+1) v_i v_m}{v_m + 2(k+1) v_i},
        &
        v_i < -\sqrt{ \frac{g x_i}{2 k(k+1)} }
        \\
        \frac{2k}{g} \frac{{v_i}^2 + {v_m}^2 + 2k v_i v_m}{v_m + 2k v_i}
        &
        \text{otherwise}.
        \end{cases}
        ,
        \qquad
        k=1,2,3,\dotsc
    .
    \end{align}
Since all paths pass through the initial point $(0,x_i)$, this point is considered a focus: $t^F_0=0$.
                
\myheading{Counting foci.}
We calculated $f(t)$, $\{f_k\}$, and $\{t^F_k\}$ explicitly in order to plot  Figure~\ref{PathDivergenceFunction}.  However, calculating the propagator only requires the Morse index $m$, which is the number of foci traversed by the path from $t=0$ to $t=T$.  This counting problem is surprisingly tricky.  Ultimately, we find that the following algorithm gives the right answer for all $(x_i,v_i,T)$:
    \begin{align}
    v_m &= \sqrt{{v_i}^2 + 2gx_i}  ,\\
    j &= \left\lfloor  \frac{gT-v_i}{2v_m} \right\rfloor  + \Theta(v_i)  ,\\
    k &= j + \Theta \left( -\sqrt{gx_i} - \sqrt{2j(j+1)} v_i  \right)  ,\\
    m^s &= j + \Theta \left( T -
        \frac{2k}{g} \frac{{v_i}^2 + {v_m}^2 + 2k v_i v_m}{v_m + 2k v_i}  
        \right)
    .
    \label{MorseIndex2}
    \end{align}
The superscript $s$ on $m^s$ is a reminder that Eq.~\eqref{MorseIndex2} pertains to the \SB{}.

\subsection{Propagator}
We are finally ready to calculate the propagator of the \SB, given the parameters $x^f_s$, $x^i_s$, and $T$.
If $x_f^s x_i^s > 0$, loop over even values of $n$ within the range defined by Eq.~\eqref{BoundOnBounces}.  Otherwise, loop over odd values of $n$.  
Use Eq.~\eqref{BackwardMapping} to obtain the parameters of the associated \QB: 
$x_i = \abs{x_i^s}$ and $x_f = (-1)^n  x_f^s  \sgn x_i^s$.
Find the allowed values of the \QB{} initial velocity, $v_{i\alpha}$, by solving Eq.~\eqref{Quartic}.  
(The \SB{} initial velocity of the \SB{} problem may be obtained as $v_i^s = v_i \sgn x_i^s$.)
Find the action $S_\alpha$, van Vleck determinant $D_\alpha$, and Morse index $m^s_\alpha$ using Eqs.~\eqref{SFromVelocitiesFull}, \eqref{VVD}, and \eqref{MorseIndex2}.
Finally, find the propagator as a sum over all classical paths:
    \begin{align}
    K^s_\text{SCA} (x_f,x_i,T) 
    &= 
    \sqrt{\tfrac{i}{2\pi\hbar}}
    \sum_{\alpha}
    \sqrt{ \abs{ D_\alpha }}
    \exp i \left( 
        \frac{S_\alpha}{\hbar} - \frac{\pi m^s_\alpha}{2} \right).
    \end{align}

\subsection{Visualizing the Semiclassical Approximation}
\begin{figure}[htbp]
    \includegraphics[width=\textwidth]{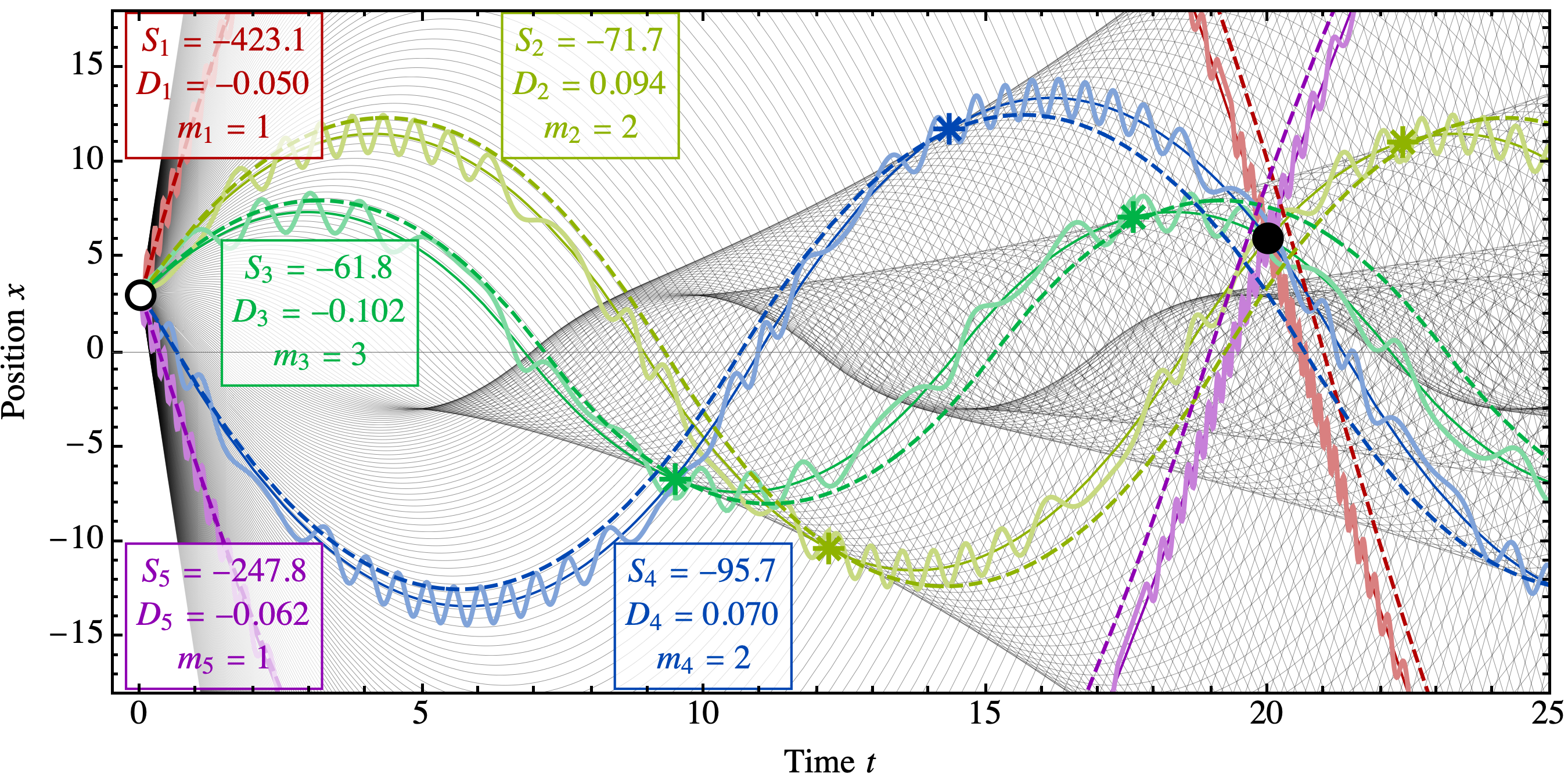}
    \caption{
    \label{Bouncer2Everything}
    \textbf{Visualizing the semiclassical approximation
    for the \SB{}.} 
    Refer to caption of Fig.~\ref{Bouncer1Everything} for meanings of symbols.
    Parameters are $x_i=3$, $x_f=6$, $T=20$, and $M=g=\hbar=1$.
    The total number of classical paths is $N_c=5$.
    }
	\end{figure}

Figure~\ref{Bouncer2Everything} visualizes the \PI{}-SCA for a \SB{} of mass $M=1$ in the potential $V(x)=Mg\abs{x}$.
The initial position is $x_i=3$.
The gray curves are classical paths $X(t)$ starting at initial position $X(0)=x_i$ (empty circle) and with a range of initial velocities $v_i \in [-5,5]$.
Each classical path has a quantum phase $\theta_\alpha(t) = S_\alpha(t)/\hbar$, visualized as cosine wiggles in the figure.
The paths overlap strongly along certain curves, called caustics.
In the semiclassical approximation, the propagator $K(x_f,x_i,T)$ is a sum of contributions from classical paths $X_\alpha(t)$ such that 
$X_\alpha(0)=x_i$ and $X_\alpha(T)=x_f$.
In this example the final point (black circle) is at $x_f=6$ and $T=20$.  There are $N_c=5$ classical paths, indicated by the colored curves.  The action $S_\alpha$ and van Vleck determinant $D_\alpha(T)$ are shown for each path.
The dashed curves result from perturbing the initial velocities by $\delta v_i=0.05$.  The stars (*) indicate foci, where paths with slightly different initial velocities converge together at a point.  The Morse index $m_\alpha$ is the number of foci along the path, where the the initial point is counted as a focus.  Each focus changes the phase by $-\pi/2$.  At each focus, the path is tangent to a caustic.  In fact, \emph{the caustics are the loci of the foci}.
The van Vleck determinant $D_\alpha$ describes the density of paths.  It is large if $x_f$ is close to a focus of path $\alpha$.

\subsection{Comparison between \PI{}-SCA and EE}
\begin{figure*}[htbp]
    \includegraphics[width=1\textwidth]{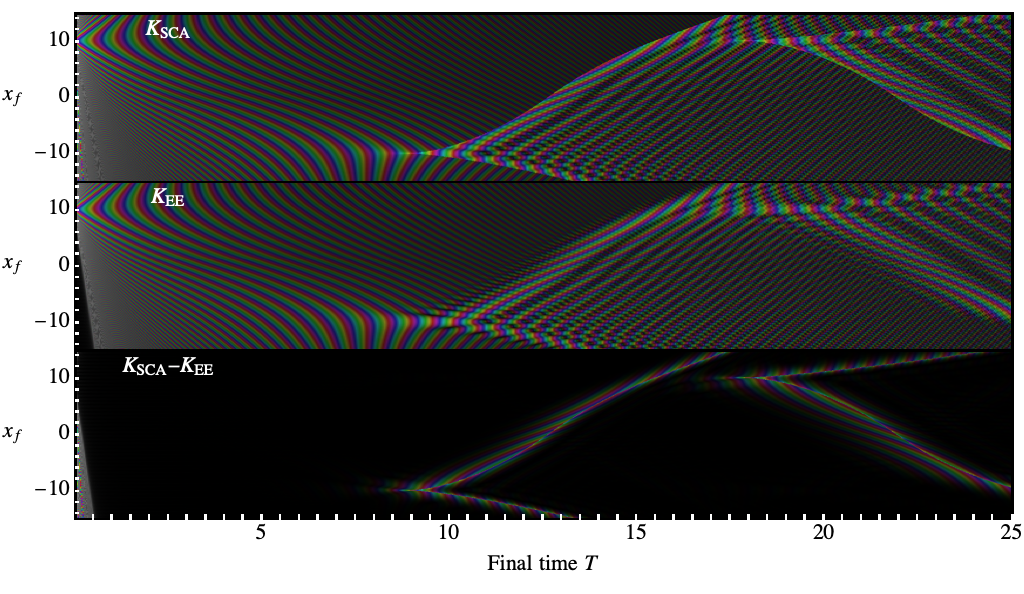}
    \caption{
    \label{Bouncer2Heatmaps}
    Propagator $K^s(x_f,x_i,T)$ for \SB{} for $x_i=10$.
    (Top) Propagator from \PI{} in semiclassical approximation, 
        $K_\text{SCA}$, shown as heatmap with rainbow palette 
        for complex magnitude and argument.
    (Center) Propagator from EE method,
	       calculated using EE truncated to $10^4$ Airy functions.  
    (Bottom) Discrepancy $K_\text{SCA} - K_\text{EE}$.
    }	      
	\end{figure*}

\begin{figure}[htbp]
    \includegraphics[width=\textwidth]{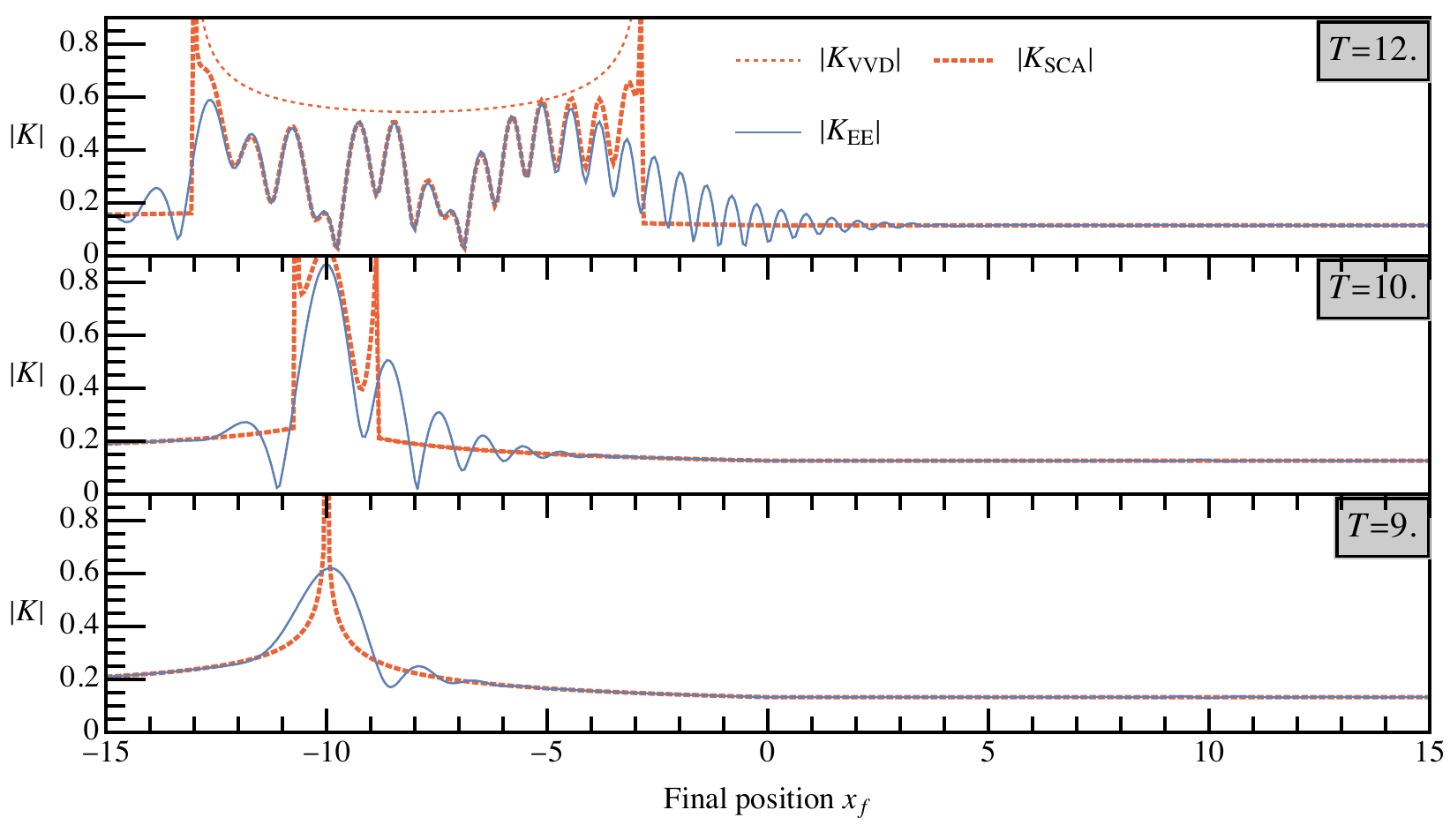}
    \caption{
    \label{Bouncer2Slices}
    Propagator $K^s(x_f,x_i,T)$ for \SB{} for $x_i=10$ at various times $T$.  
    $K_\text{SCA}$ is the propagator in the semiclassical approximation.
    $K_\text{VVD}$ is the envelope obtained by ignoring phases.
    $K_\text{EE}$ is the propagator from the EE method.
    }
	\end{figure}

Figures \ref{Bouncer2Heatmaps} and \ref{Bouncer2Slices} show the propagator calculated using the \PI{}-SCA and EE.  As before, the agreement is excellent except near the caustics.  The discussion of the results is similar to what we have presented for the \QB{}.

Remarkably, the features in the propagator emerge from the EE calculation without explicit consideration of classical paths.  Those features are encoded in the Airy functions and Airy zeroes, but in a way that is difficult if not  impossible for a human to extract.  The great value of the \PI{}-SCA approach is that it explains all the features in terms of caustics.

\section{Goodman Subtraction}  \label{secGoodmanSubtraction}
In this section we extend Goodman's argument\cite{Goodman81v49} to show that a constrained path integral can be replaced by a subtraction between two unconstrained path integrals.

\begin{figure}[htbp]
    \includegraphics[width=0.49\textwidth]{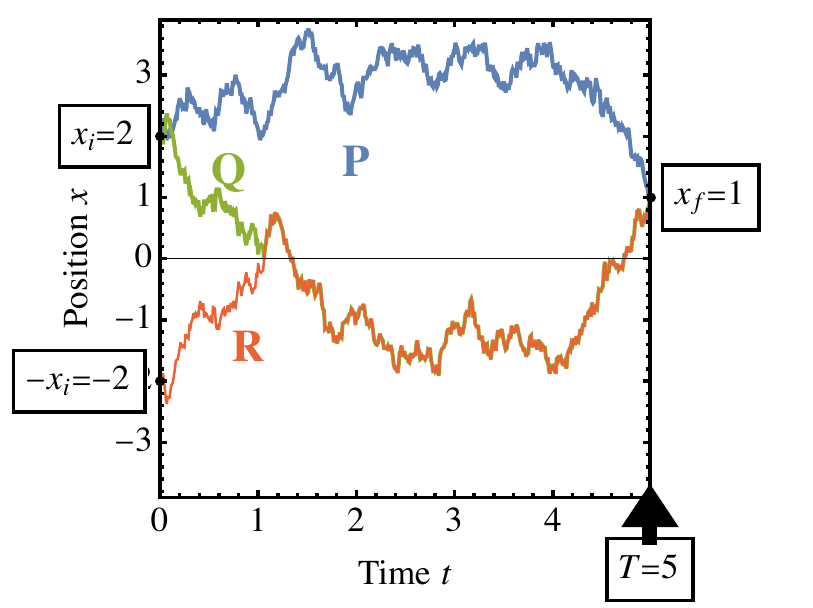}
    \caption{
    \label{figGoodmanSubtraction}
    \textbf{(Color online)}
    Visualizing the Goodman subtraction method.
    }
	\end{figure}

\myheading{Hard wall.}
Consider a quantum-mechanical particle near a hard wall such that $V(x)=\infty$ if $x<0$ and $V(x)=0$ if $x<0$.  We will use the shorthand notation $V(x) = \infty~ \Theta(-x)$.  
This problem can be solved using the method of images\cite{Schulman2005-book}, for which a rigorous derivation has been given by Goodman\cite{Goodman81v49}.  Goodman's subtraction argument is visualized in Fig.~\ref{figGoodmanSubtraction}.

Let P be the set of valid paths, $x(t)$, which stay in the allowed region ($x>0$) for all times $t$.
The propagator for a particle near a hard wall is an integral over all paths in P:
$K_{{\rm hw}} (x_f, x_i, T) = K_P \equiv \int_P \scD x~ \exp iS[x]/\hbar$.
The green curve in the figure shows one representative path from set P.

Let Q be the set of invalid paths that enter the forbidden region ($x<0$).
The free-particle propagator from the initial position to the final position is a sum over both valid paths (P) and invalid paths (Q):
$K_0(x_f,x_i,T) = K_P + K_Q$.

For any path $x(t)$ in Q, $x=0$ at one or more time points $(t_1, t_2, \dotsc)$.  
Construct an image path $x_\text{image} (t) = [2\Theta(t-t_1) - 1] x(t)$ by reflecting the segment of the path before the first zero-crossing, as shown in the figure. 
Let R be the set of image paths constructed by applying this transformation to every path in Q.
The transformation introduces no discontinuities into $x_\text{image} (t)$,
so $\dot{x}_\text{image} (t) = \dot{x} (t)$ for all $t$,
and the free-particle action $\int_0^T dt~ \frac{M}{2} \dot{x}^2$ is unchanged.

The free-particle propagator from the \emph{image} of the initial position, $-x_i$, to the final position, is a sum over paths in R only:
$K_0(x_f,-x_i,T) = K_R$.
The free-particle action is invariant under the partial reflection transformation, so $K_Q = K_R$.
Since $(K_P+K_Q)-(K_R) = K_P$, we may replace the \emph{constrained path integral} by the \emph{difference of two unconstrained path integrals}:
    \begin{align}
    K_\text{hw} (x_f,x_i,T)
    &=  K_0(x_f,x_i,T)
    -   K_0(x_f,-x_i,T)
    .
    \label{KHardWall}
    \end{align}

\myheading{Bouncer.}
Now consider the \QB{}, whose potential include both gravity and an impenetrable ground as depicted in Fig.~\ref{VBouncer}.  The propagator is
    \begin{align}
    K(x_f,x_i,T)
    &=\int_{x(0)=x_i}^{x(T)=x_f}  \scD x~ 
        \exp \frac{i}{\hbar}
        \int_0^T dt~ 
        \left[ \frac{M}{2} \dot{x}^2 - Mgx - \infty~ \Theta(-x) \right].
    \label{KBouncerPI}
    \end{align}
We may replace the infinite potential by a restriction on the domain of integration to the region $x>0$.  We may then redefine $V(x)$ for $x<0$ in any convenient way.  For reasons that will become clear later, we choose to set $V(x)=-x$ for $x<0$.  Then we may replace $x$ by $\abs{x}$ in the exponent:
   \begin{align}
    K(x_f,x_i,T)
    &=\int_{x(0)=x_i, x(t)>0 \forall t}^{x(T)=x_f}  \scD x~ 
        \exp \frac{i}{\hbar}
        \int_0^T dt~ 
        \left( \frac{M}{2} \dot{x}^2 - Mg\abs{x} \right)
    \end{align}

We now extend Goodman's original argument as follows, referring again to Fig.~\ref{figGoodmanSubtraction}.
As before, consider any path $x(t)$ that touches or crosses zero.  Perform the partial reflection, $x_\text{image} (t) = [2\Theta(t-t_1) - 1] x(t)$.  The free-particle action,
$\int_0^T dt~ \frac{M}{2} \dot{x}^2$, is unchanged.
The time integral of the symmetrized potential, $\int_0^T dt~ Mg\abs{x}$, is also unchanged.
Therefore we may again replace the \emph{constrained path integral} by the \emph{difference of two unconstrained path integrals}:
   \begin{align}
    K(x_f,x_i,T)
    &=
    K_s (x_f,x_i,T) - K_s (x_f,-x_i,T)
    ,
    \label{KBouncerSubtractionMethod}
    \end{align}
where
   \begin{align}
    K_s (x_f,x_i,T)
    &=
    \int_{x(0)=x_i}^{x(T)=x_f}  \scD x~ 
        \exp
        \frac{i}{\hbar}
        \int_0^T dt~ 
        \left( \frac{M}{2} \dot{x}^2 - Mg\abs{x} \right)
        .
    \label{Ks}
    \end{align}
is the propagator for the \SB{} with $V_s(x)=Mg\abs{x}$.

The infinite potential ramp $Mgx$ and the infinite potential barrier $\infty\,\Theta(-x)$ do not individually pose serious problems: $K_0$, $K_g$, and $K_\text{hw}$ can all be found in closed form.  However, the combination of these potentials leads to a problem of much greater difficulty.
To use Goodman's path cancellation argument to replace the barrier by an image source, we must symmetrize the ramp potential, leading to a non-analytic function $V_s=Mg\abs{x}$. 
Because of this, Eq.~\eqref{Ks} is a non-Gaussian path integral that cannot be evaluated exactly.  

In Sec.~\ref{secBouncer2} we presented the calculation of Eq.~\eqref{Ks} within the semiclassical approximation.  In Eq.~\eqref{KBouncerSubtractionMethod}, the first \SB{} propagator $K_s(x_f,x_i,T)$ has initial and final positions of the same sign, so it involves only paths with an even number of bounces.  In contrast, the second term  $K_s(x_f,-x_i,T)$ has initial and final positions of opposite signs, so it involves only paths with an odd number of bounces.  Therefore, the \QB{} propagator $K(x_f,x_i,T)$ contains contributions from paths with any number of bounces ($n=0,1,2,3,\dotsc$), but these contributions are weighted by a factor of $(-1)^n$.  In other words, each bounce causes a phase change of $\pi$.

Recall that in the SCA formula, each focus causes a phase change of $-\frac{\pi}{2}$.  Thus each bounce is equivalent to two additional foci.  The Morse index for a path of the \QB{} is $m = m_s + 2n$.

    \begin{figure*}[!htbp]
        \includegraphics[width=1\textwidth]{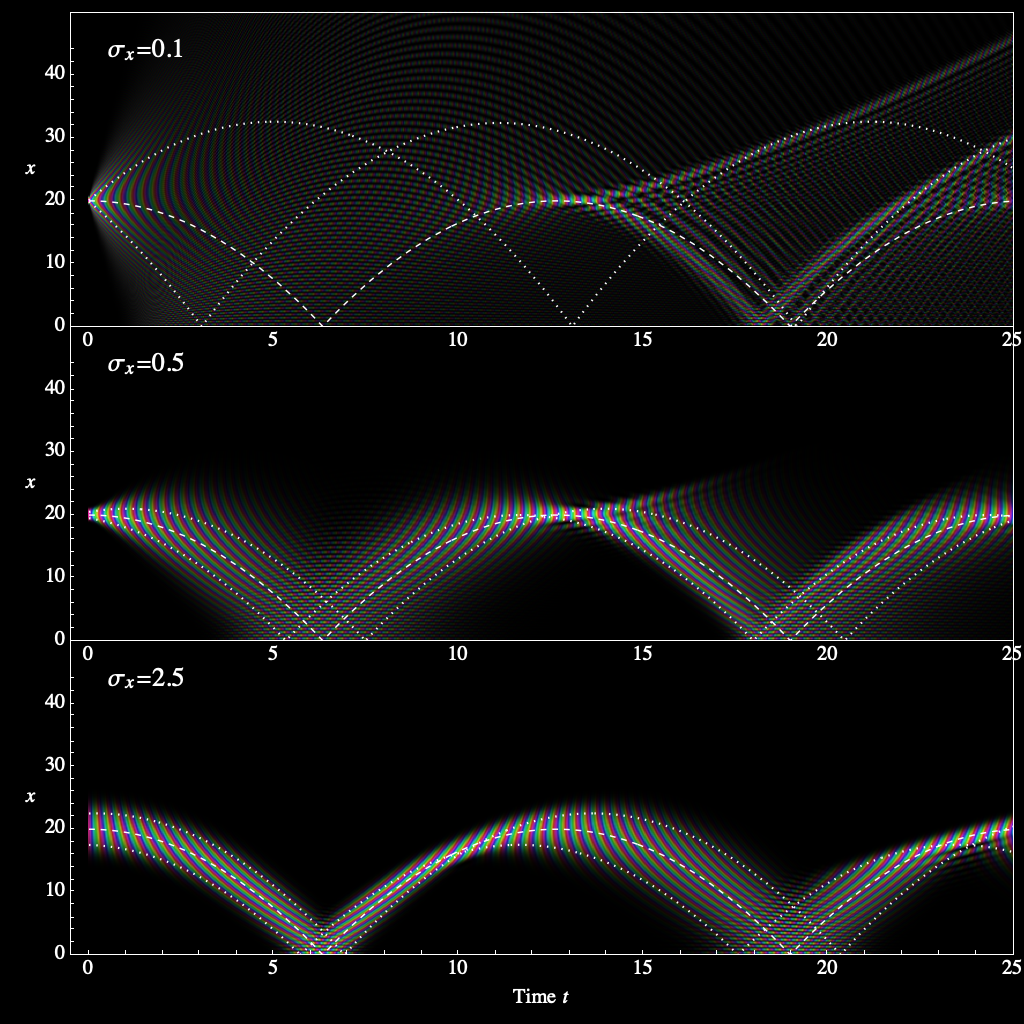}
    \caption{
	    \label{WavePacketEvolution}
        \textbf{Evolution of three minimum-uncertainty Gaussian wave packets in the \QB.}
        Rainbow colors show wave function $\Psi(x,t)$
        evolved using EE truncated to $10^4$ Airy functions.   
	    Dashed curves indicate classical paths with initial positions
	    $x(0) = x_i \pm \sigma_x$ and initial momenta
	    $p(0)= p_i \pm \sigma_p$
	    where $\sigma_x$ and $\sigma_p$ are the position and momentum uncertainties of the wave packet.
    }
	\end{figure*}

\section{Comparison with Wave Packet Evolution} \label{secWavePacketEvolution}
\myheading{Quantum revivals.}
It is instructive to compare the propagator with the time evolution of a Gaussian wave packet for the \QB{}. Here we visualize the time-dependent wave function $\Psi(x,t)$ itself.  This can be calculated in the standard way using the EE:
    \begin{align}
    c_n(0) &= \int_0^\infty dx~ \varphi^\star_n(x) \Psi(x,0)
        ,\nonumber\\
    c_n(T) &= e^{i\lambda_n T/\gamma T_0} c_n(0)
        ,\nonumber\\
    \Psi(x,T) &= \sum_{n=1}^\infty \varphi_n(x) c_n(T).
    \end{align}
For efficient calculation, we discretize the integral, truncate the sum, and cast the equations into matrix-vector form.  Figure~\ref{WavePacketEvolution} shows the time evolution of Gaussian wave packets of the form
    \begin{align}
    \Psi(x,0) &= (2\pi)^{-1/4} {\sigma_x}^{-1/2} 
    e^{   -(x-x_i)^2/4{\sigma_x}^2  } 
    \end{align}
with initial position $x_i=20$. 
In the bottom panel, the initial wave packet is well localized in real space ($\sigma_x=2.5$) and in momentum space ($\sigma_p=1/2\sigma_x=0.2)$.  As expected, the wave function initially follows a classical multiple-bounce trajectory.  Later on, it spreads out due to dephasing between the components with different initial positions and momenta.  This type of situation has been studied in detail by Gea-Banacloche\cite{Gea-Banacloche99v67}, who found that the mean altitude $\mean{x(t)}$ undergoes oscillations, and the amplitude of these oscillations exhibits collapses and partial revivals.  The phenomenon of \emph{quantum revivals}\cite{eberly1980,gaeta1990} is usually associated with \emph{commensurate differences between eigenfrequencies}.
In the top panel, the initial wave packet is almost a Dirac delta function in real space ($\sigma_x=0.1$), but it has a large momentum uncertainty ($\sigma_p=5$).  At first the wave packet disperses strongly, so that the probability weight becomes spread out over a wide range of $x$, but the probability density partially re-converges at the beginning of the first caustic.  Remarkably, the particle seems to disappear at short times ($t\sim 5$) but reappears near the initial position ($x_i=20$) at $t \approx 13.5$, and then splits into two!  We conjecture that this is a \emph{different type of quantum revival that is associated with caustics in spacetime}.  

Caustic-based quantum revivals may be experimentally observable if the coherence time is long enough.  Possible experimental realizations include the following.

\myheading{Ultracold atoms.}
Trapped ultracold atomic gases provide a versatile platform for realizing many quantum-mechanical models. Neutral atoms experience a potential due to applied static and dynamic electric and magnetic fields, which may be supplied by plates, coils, wires,\cite{pietra2005} microwaves, or lasers. A common technique in cold atom experiments is to confine an atom cloud in a narrow trapping potential, then suddenly switch off the narrow potential and allow the cloud to expand in a potential with a larger length scale.\cite{pietra2005,liao2010,sommer2011,greiner2002} This approach can certainly be applied to the \QB.

\myheading{Magnons.}
When a spatially varying applied magnetic field $B(x)$ is applied to a ferromagnetic nanowire, the magnons (quantized spin wave excitations) have a dispersion relation of the form $\omega(k) = \alpha B(x) + \hbar^2 k^2 / 2M$, where $\alpha$ is some constant and $M$ is the effective mass of the magnons. In this system, a magnon behaves identically to a particle moving in a 1D potential. One may excite a localized magnon and study how it spreads, using optical, spintronic, or scanning SQUID methods with sufficient spatial and temporal precision.

\myheading{Electrons.}
If an electron is suddenly introduced into a semiconducting nanowire with a lengthwise electric field, it might be expected to behave as a quantum bouncer. However, such an experiment is likely to be complicated due to electron-electron interaction, impurity scattering, and the difficulty of injecting an electron at a definite time.

\section{Discussion and Conclusions} \label{secConclusions}
For the \QB{} and \SB{}, the number of classical paths is always finite.  This suggests, ironically, that the \PI{}-SCA for the quantum bouncers is ``simpler'' than the EE method (which requires an infinite sum), and that the quantum bouncers are  ``simpler'' than the particle-in-a-box or particle-on-a-ring problems, where both \PI{} and EE methods require summing over an infinite number of paths or states!  Of course, the \PI{}-SCA is merely an approximation.  To make it exact would require summing infinitely many higher-order terms coming from the power series expansion of the action beyond second order in fluctuations, which is difficult and computationally costly.

Near the caustics, the Gaussian approximation breaks down, and Eq.~\eqref{KSemiclassical} is inaccurate.  It is possible to salvage the SCA by including higher-order terms in the expansion around the classical paths.\cite{Schulman2005-book}  This is beyond the scope of this paper.

In conclusion, in this paper we have used the Feynman path integral (\PI{}) formalism with a semiclassical approximation (SCA) to evaluate the propagators of the one-sided bouncer (\QB{}) and the symmetric bouncer (\SB{}).  We have showed how an infinite potential barrier can be treated using the method of images.   We have developed visualization methods for the \PI{} that are appealing yet quantitative.  We have verified that the \PI{}-SCA method agrees with the eigenfunction expansion (EE) method for most parameter combinations.  
The path integral formalism explains how the features in heatmaps of the propagator are related to the caustics, which are phase boundaries at which the number of classical paths changes (these features are difficult to explain from the EE point of view).
We point out that the quantum bouncer exhibits quantum revivals that are associated with caustics in spacetime (rather than commensurate eigenfrequencies).
We suggest various experimental realizations of our model and implications of the results.

\emph{Mathematica} notebooks with interactive visualizations are provided as Supplementary Information,
and on GitHub at \url{https://github.com/lohyenlee/quantum-bouncer}.  


\end{document}